\documentclass[aps,pra,twocolumn,nopacs,showpacs,superscriptaddress]{revtex4-1}

\usepackage{graphicx}  
\usepackage{dcolumn}   
\usepackage{bm}        
\usepackage{amssymb}   
\usepackage{amsmath}
\usepackage{units}
\usepackage[ansinew]{inputenc}
\usepackage[dvipsnames]{xcolor}
\usepackage{lipsum}
\usepackage{ulem}

\usepackage[type1]{libertine}                                        
\usepackage{textcomp}
\usepackage[scaled=.85]{beramono}
\usepackage[libertine,cmintegrals,cmbraces,vvarbb,slantedGreek]{newtxmath}
\usepackage[scr=boondoxo]{mathalfa}
\usepackage{bm}
\usepackage[lf]{carlito}

\usepackage{braket}
\usepackage{bm}

\makeatletter
\DeclareRobustCommand{\cev}[1]{%
  \mathpalette\do@cev{#1}%
}
\newcommand{\do@cev}[2]{%
  \fix@cev{#1}{+}%
  \reflectbox{$\m@th#1\vec{\reflectbox{$\fix@cev{#1}{-}\m@th#1#2\fix@cev{#1}{+}$}}$}%
  \fix@cev{#1}{-}%
}
\newcommand{\fix@cev}[2]{%
  \ifx#1\displaystyle
    \mkern#23mu
  \else
    \ifx#1\textstyle
      \mkern#23mu
    \else
      \ifx#1\scriptstyle
        \mkern#22mu
      \else
        \mkern#22mu
      \fi
    \fi
  \fi
}
\makeatother

\newcommand{\figref}[1]{Fig.\,\ref{#1}}
\renewcommand{\eqref}[1]{Eq.\,(\ref{#1})}
\newcommand{\suppcite}[1]{\cite[S\hspace{-1mm}][]{#1}}

\renewcommand{\Re}{\text{Re}}

\begin{document}

\title{Microwave-assisted tunneling and interference effects in superconducting junctions under fast driving signals}

\author{Piotr Kot}
\affiliation{Max-Planck-Institut f\"ur Festk\"orperforschung, Heisenbergstra{\ss}e 1,
70569 Stuttgart, Germany}
\author{Robert Drost}
\email[Corresponding author; electronic address:\ ]{r.drost@fkf.mpg.de}
\affiliation{Max-Planck-Institut f\"ur Festk\"orperforschung, Heisenbergstra{\ss}e 1,
70569 Stuttgart, Germany}
\author{Maximilian Uhl}
\affiliation{Max-Planck-Institut f\"ur Festk\"orperforschung, Heisenbergstra{\ss}e 1,
70569 Stuttgart, Germany}
\author{Joachim Ankerhold}
\affiliation{Institut f\"ur Komplexe Quantensysteme and IQST, Universit\"at Ulm, Albert-Einstein-Allee 11, 89069 Ulm, Germany}
\author{Juan Carlos Cuevas}
\affiliation{Departamento de F\'{\i}sica Te\'orica de la Materia Condensada and
Condensed Matter Physics Center (IFIMAC), Universidad Aut\'onoma de Madrid, 28049 Madrid, Spain}
\author{Christian R. Ast}
\affiliation{Max-Planck-Institut f\"ur Festk\"orperforschung, Heisenbergstra{\ss}e 1,
70569 Stuttgart, Germany}

\date{\today}

\begin{abstract}

As scanning tunneling microscopy is pushed towards fast local dynamics, a quantitative understanding of tunnel junctions under the influence of a fast AC driving signal is required, especially at the ultra-low temperatures relevant to spin dynamics and correlated electron states. We subject a superconductor-insulator-superconductor junction to a microwave signal from an antenna mounted \textit{in situ} and examine the DC response of the contact to this driving signal. Quasi-particle tunneling and the Josephson effect can be interpreted in the framework of Tien-Gordon theory. The situation is more complex when it comes to higher order effects such as multiple Andreev reflections. Microwave assisted tunneling unravel these complex processes, providing deeper insights into tunneling than are available in a pure DC measurement.

\end{abstract}


\maketitle

\section{Introduction}

With its combination of supreme spatial resolution and spectroscopic imaging, the scanning tunneling microscope (STM) is a workhorse of mesoscopic and nanoscale physics\,\cite{binnig1982surface}. The technique remains plagued, however, by the inherently low bandwidth of the transimpedance amplifiers required to measure the small tunnel current. Even so, there is a growing desire to exploit the unique capabilities of the STM to study nanoscale objects on their own time scales. With the integration of high-frequency excitation methods and development of pump-probe schemes in recent years, this goal appears to be in reach\,\cite{loth2010measurement, terada2010laser, cocker2013ultrafast, yoshida2014probing, rashidi2016time, cocker2016tracking, garg2019attosecond}. A fast driving signal is generally viewed as a means to excite fundamental modes within the sample and study their dynamics. The effect of the AC signal on the DC conductance is much less well explored. In this context, a quantitative understanding of the response of the tunnel junction itself to a fast driving signal is required, especially at the ultra-low temperatures relevant to spin dynamics and correlated electron states.

Previous theoretical and experimental investigations into this direction have focussed on tunneling between superconductors where conductance spectra are dominated by sharp peaks which make the effect of microwave radiation easy to discern\,\cite{shapiro1963josephson, tien1963multiphoton, falci1991quasiparticle, chauvin2006superconducting, roychowdhury2015microwave}. This situation is especially interesting as superconductor-insulator-superconductor (SIS) junctions support multiple Andreev reflections (MARs), permitting us to study the interaction of microwave radiation with higher order tunneling processes. On the theoretical level, the problem is treated semi-classically with the microwave signal leading to a time-dependent modulation of the bias voltage dropping across the junction. Quasi-particles (QPs) and Cooper pairs (CPs) thus show a similar reaction to the incident radiation despite their stark differences in physical origin. It has been suggested that microwave radiation couples to the tunnel current through the total charge transferred between the electrodes in any given process\,\cite{roychowdhury2015microwave}.

\begin{figure}
	\centering
		\includegraphics[width=1.00\columnwidth]{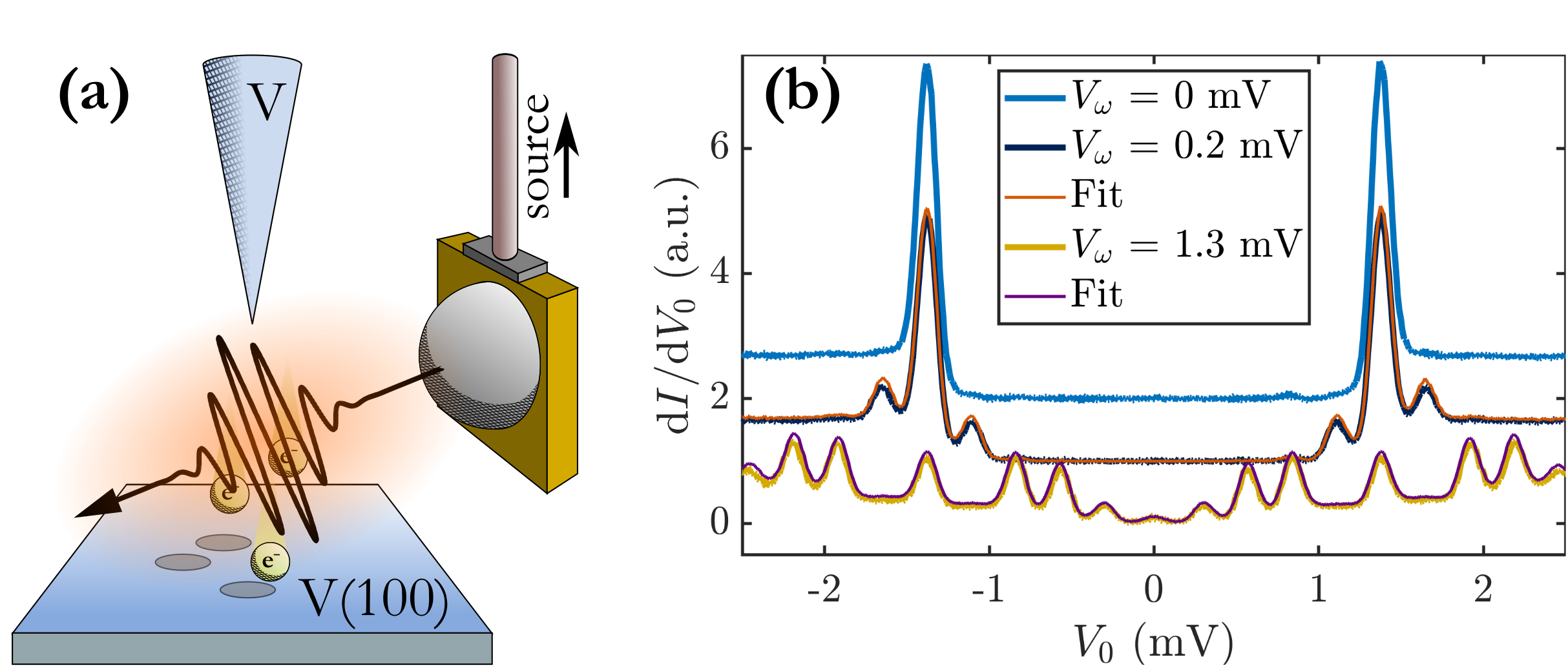}
		\caption{Microwave assisted tunneling in an SIS junction: (a) Sketch of the experimental setup. A commercial STM is fitted with a radio frequency antenna delivering radiation to the junction. (b)  Conductance spectrum in the tunneling regime without microwave signal (light blue) and under microwave radiation at two different power settings (dark blue with $V_{\omega} =\,$0.2\,mV and yellow with $V_{\omega} =\,$1.3\,mV) with corresponding fits according to \eqref{eq:tiengordon} respectively in orange and purple. $G_N \approx 8.4\times10^{-3}\,G_0$ for all spectra.}
	\label{fig:figure1}
\end{figure}

Most data up to date has been analyzed in the framework of the Tien-Gordon model, which treats tunneling and microwave interaction as independent processes\,\cite{tien1963multiphoton, falci1991quasiparticle}. More advanced theories suggest that SIS tunneling under microwave irradiation must be understood through MARs while accounting for microwave interactions at every step of these higher order processes\,\cite{cuevas2002subharmonic}. In this MAR model, the microwave signal is modelled as a time-dependent phase difference between tip and sample electrodes and MARs arise naturally through higher order terms. Interference between particles traveling back and forth within the barrier are predicted to yield a very different result than the Tien-Gordon approach. Here, we report measurements of QP, CP, and MAR tunneling in an SIS junction under microwave irradiation. These experimental findings are compared with predictions from the Tien-Gordon and the microwave-assisted MAR models, respectively. Only the latter one is found to be capable of correctly modeling the impact of microwave driving on the charge transfer process.

\section{Experimental}

We study SIS tunneling between a V(100) surface and a vanadium tip in a commercial STM system (Unisoku USM1300) operating at 300\,mK base temperature and fitted with a custom-built antenna assembly capable of delivering a microwave signal between 60 to 90\,GHz to the junction, see \figref{fig:figure1}(a). The V(100) sample is prepared by repeated bombardment with Ar$^+$ ions at 1\,keV and annealing at $\approx$650\,$^{\circ}$C, resulting in the well-known (5$\times$1) oxygen reconstruction\,\cite{davies1980surface, foord1983the}. The tip is cut from vanadium wire and cleaned by bombardment with Ar$^+$ ions before being transferred into the STM and further prepared by controlled indentation into the V(100) surface until a clean SIS signature is observed.

\section{Measurement and Discussion}

\begin{figure}
	\centering
		\includegraphics[width=1.00\columnwidth]{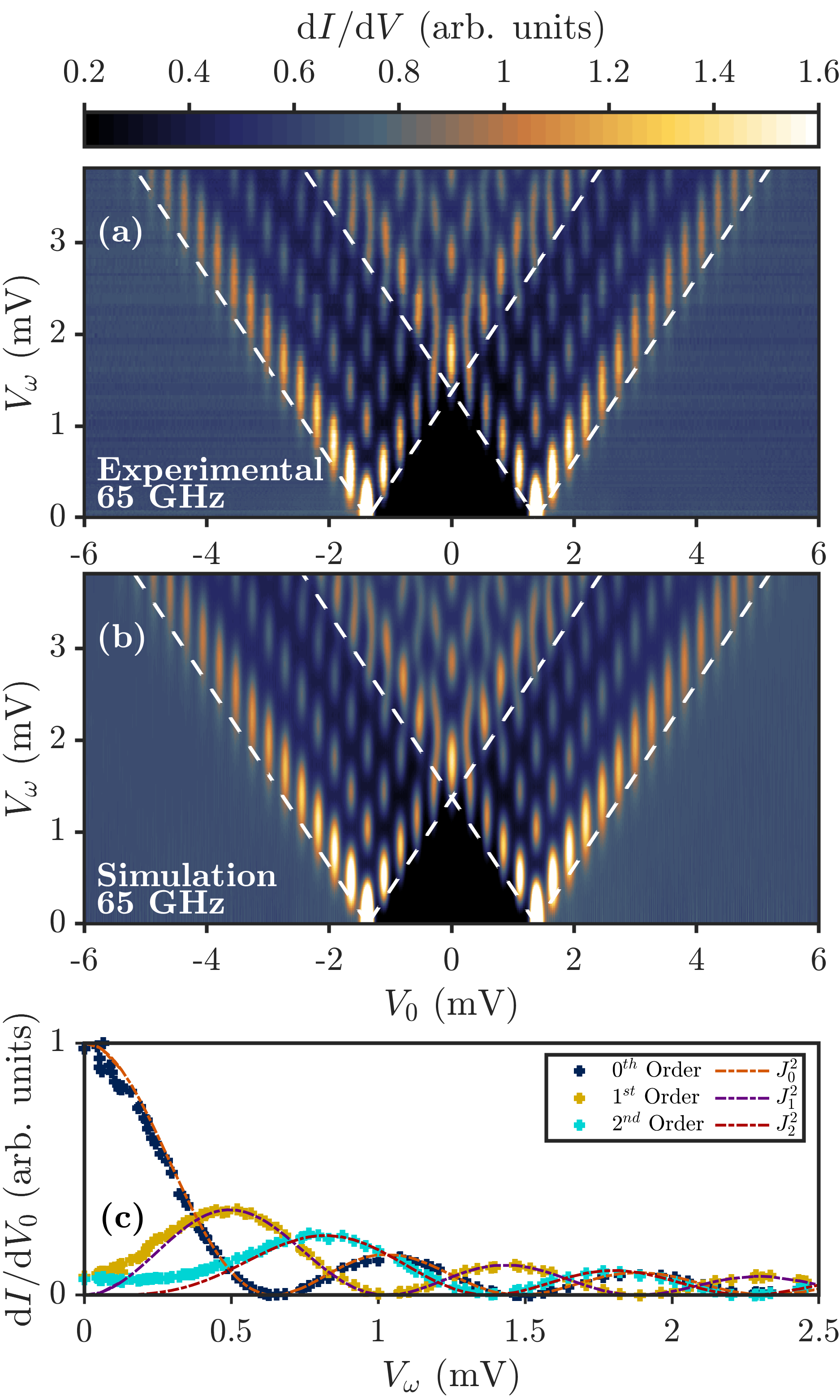}
	\caption{$V_{\omega}$ dependence of microwave assisted tunneling: (a) Stacked contour plot of a series of conductance spectra of an SIS junction under irradiation by a 65\,GHz signal at varying $V_{\omega}$. The white dashed lines demark the energy of the coherence peaks $\pm V_{\omega}$. $G_N = 8.4\times10^{-3}\,G_0$. (b) Simulation of the data shown in panel (a) using \eqref{eq:tiengordon}. (c) Constant $V_{0}$ profiles of the experimental data in panel (a) at the principal coherence peak (dark blue) and the first (yellow) and second (cyan) satellites, normalised to the maximum of the $0^{th}$ order peak. The dashed-dotted lines are the Bessel functions of zeroth (orange), first (purple), and second (red) order in $\frac{e V_{\omega}}{\hbar \omega}$.}
	\label{fig:figure2}
\end{figure}

We begin our investigation in the tunneling regime at large tip-sample distances where the normal state conductance $G_N$ is only a small fraction of the quantum of conductance $G_0 = 2e^2/h$, where $e$ is the elementary charge and $h$ the Planck constant. Conductance spectra acquired above the pristine V(100) surface show behavior typical for SIS junctions. A representative data set is shown in \figref{fig:figure1}(b) in light blue. The primary features are two sharp coherence peaks separated by twice the sum of the superconducting gap in tip and sample, $\Delta_{\mathrm{tip}}$ = 663\,$\mu e$V and $\Delta_{\mathrm{sample}}$ = 710\,$\mu e$V (see Supplemental Material for details)\,\cite{si}. Such a low-temperature, low-capacitance junction under AC driving has previously been considered by Falci\,\textit{et al.}, who found the following expression for the time-averaged DC tunnel current, also known as the Tien-Gordon equation\,\cite{tien1963multiphoton, falci1991quasiparticle, si}:

\begin{figure*}[!ht]
	\centering
		\includegraphics[width=1.00\textwidth]{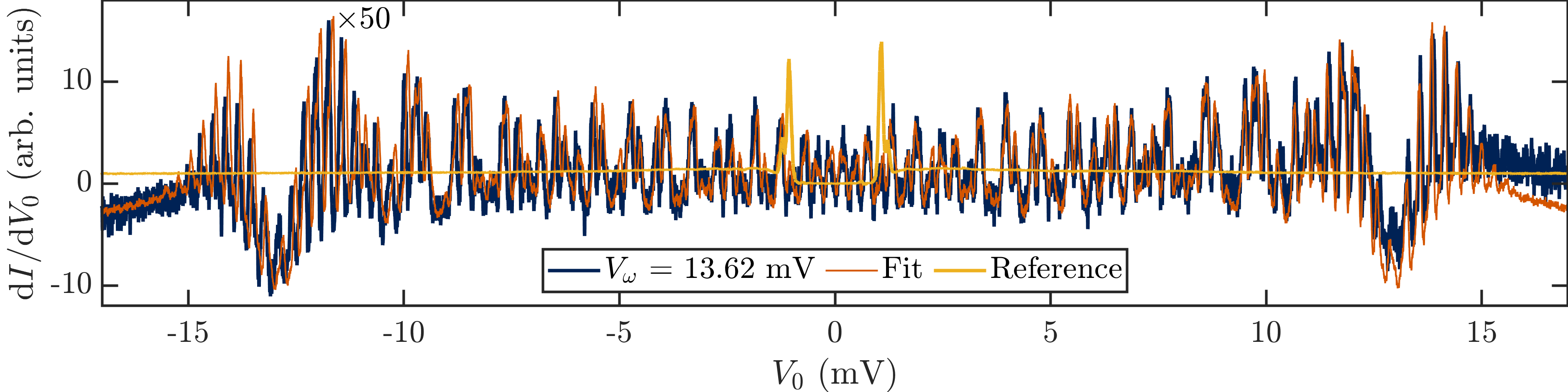}
	\caption{Junction at high microwave amplitude: Conductance spectrum of an unperturbed SIS junction (yellow) and of the same junction under irradiation by the maximum possible RF signal in our setup (dark blue), equivalent to $V_{\omega} = 13.62$\,meV at 70.02\,GHz. The orange curve is a fit to the data using \eqref{eq:tiengordon} and the reference spectrum in yellow as input data. Even at these extreme conditions, the data is well reproduced. Discrepancies on the high positive bias side are due to a slight and unavoidable z-drift during data acquisition. The in-gap peaks in the reference spectrum arise from a magnetic impurity state on the tip. $G_N = 5\times10^{-3}\,G_0$.}
	\label{fig:figure3}
\end{figure*}

\begin{equation}
    I(V_{0}, V_{\omega}) = \sum^\infty_{n=-\infty} J_n^2 \left( \frac{e\text{V}_{\omega}}{\hbar \omega}\right ) I_0 \left( V_{0}-\frac{n \hbar \omega}{e}\right). \label{eq:tiengordon}
\end{equation}

\noindent Here, $V_{0}$ is the DC junction bias, $V_{\omega}$ the amplitude of the AC voltage resulting from the incident microwave, $J_n$ are the Bessel functions of the first kind and order $n$, $e$ is the elementary charge, $\hbar$ the reduced Planck constant, $\omega$ the angular frequency of the microwave signal, and $I_0$ is the tunnel current in absence of irradiation. The same relationship follows for the conductance signal $G(V_{0}, V_{\omega}) = \frac{\mathrm{d}}{\mathrm{d}V_{0}} I(V_{0}, V_{\omega})$. Note that though we study a SIS junction in the present case, \eqref{eq:tiengordon} will also hold for normal conducting samples.

We thus expect to observe a weighted replication of the original signal at integer multiples of $\hbar \omega$ when subjecting the junction to a microwave signal. This is indeed what is observed in previous experiments\,\cite{soerensen1974microwave, kouwenhoven1994photon, degraaf2013charge}. Data from a tunnel junction under microwave irradiation is shown in \figref{fig:figure1}(b) for two different source power settings in dark blue and yellow to highlight the effect of changing $V_{\omega}$. The initial effect of a low-power microwave signal is a diminishing of the coherence peaks at $\pm ( \Delta_{\mathrm{tip}} + \Delta_{\mathrm{sample}})$, coupled with the appearance of satellite peaks offset by $\pm \hbar \omega$ (see dark blue curve in \figref{fig:figure1}(b)\,). As the power level of the microwave source is increased, higher order replica of the coherence peaks appear at integer multiples of $\pm \hbar \omega$ (see yellow curve in \figref{fig:figure1}(b)\,). We use \eqref{eq:tiengordon} to fit the experimental data using a reference spectrum, acquired in absence of microwave radiation at the same sample location, and the known value for $\omega$, set at the source module, as input with $V_{\omega}$ as the only free parameter. Fits to the experimental data are shown in orange and purple in \figref{fig:figure1}(b).

The predicted weighting of the replica by $J_n^2 ( e\text{V}_{\omega} / \hbar \omega )$ becomes apparent when varying the source power level, and thereby $V_{\omega}$, at fixed frequency. We performed such a measurement, acquiring a series conductance curves at a frequency of 65\,GHz with amplitudes $V_{\omega}$ between 0 and 3.8\,mV. Each curve is fitted using \eqref{eq:tiengordon} to extract $V_{\omega}$. The resulting data set is shown as a stacked contour plot in \figref{fig:figure2}(a). The coherence peaks fan open as $V_{\omega}$ gradually increases and the spectral intensity of the principal peaks is distributed over a wider energy range. A rich interference pattern emerges in a widening area around zero bias as replica from above and below the Fermi level are woven together. The resulting pattern can be fully understood by the superposition of coherence peak replica. The agreement between the measured data and the model in \eqref{eq:tiengordon}, shown in \figref{fig:figure2}(b), is remarkable.

It is important to note that the energy offset of $\hbar\omega$ between the peak replica should not be interpreted in terms of photon absorption from the microwave field during the tunneling process. Instead, \eqref{eq:tiengordon} is the result of the quantum mechanical motion of the electron in a time-dependent classical potential of the form $U(t) = e V_{\omega} \cos(\omega t)$. The inclusion of such a potential will naturally lead to the emergence of the Bessel functions and the resulting eigenenergies will contain components at $E_0$, $E_0 \pm \hbar \omega$, $E_0 \pm 2 \hbar \omega$ and so on\,\cite{tien1963multiphoton}. As such, the emergence of the satellite peaks in microwave assisted tunneling are a manifestation of the 	quantum mechanical nature of electrons in solids. Indeed, the work performed by the classical microwave field on electrons surpasses the photon energy by far, thus placing the experiment in a field-driven regime\,\cite{kampfrath2013resonant, si}.

We extract profiles at constant $V_{0}$, shown in \figref{fig:figure2}(c), from the data and normalise the resulting curves to the maximum intensity of the coherence peak at $V_{\omega} = 0$. The resulting profiles follow their respective low-order Bessel functions almost exactly. Minor discrepancies arise only at low $V_{\omega}$, where the normal conducting background signal has a significant contribution.

It is interesting to note that \eqref{eq:tiengordon} gives an accurate description of the junction even as $e V_{\omega} \gg 2\Delta_{\mathrm{sample}}$ (ca. 1.52\,mV for the present case of vanadium). To push the limits of the model, we examined the behavior of a typical SIS junction when subjected to the highest intensity of microwave radiation possible in our setup, where $V_{\omega}$ reaches 13.62\,mV at 70.02\,GHz, corresponding to $h \nu = 289\,\mu$eV, and is thus nearly an order of magnitude larger than $2\Delta_{\mathrm{sample}}$. The corresponding data is shown in \figref{fig:figure3} (the in-gap peaks in the reference spectrum in \figref{fig:figure3} arise from a magnetic impurity state on the tip). Even in these extreme conditions, the experimental data is still well described by the Tien-Gordon \eqref{eq:tiengordon}. Given that an unperturbed reference spectrum is used in our modeling, this result suggests that the superconductor remains undisturbed by the microwave signal. Indeed, the superconductor is transparent at energies $\hbar \omega < 2\Delta_{\mathrm{sample}}$ as there are no final states available into which CPs could be excited\,\cite{tinkham}.

A closer look at the Josephson effect further offers the possibility to study the effect of microwave radiation on tunneling between coherent electron states. Such measurements have previously been performed by Roychowdhury and co-workers, also employing the theoretical description by Falci \textit{et al.}, who gave an expression strikingly similar to the Tien-Gordon equation for the case of the Josephson current in the tunneling regime\,\cite{falci1991quasiparticle, grabert2015dynamical, roychowdhury2015microwave}:

\begin{equation}
    I(V_{0}, V_{\omega}) = \sum^\infty_{n=-\infty} J_n^2 \left( \frac{2e\text{V}_{\omega}}{\hbar \omega}\right ) I_0 \left( V_{0}-\frac{n \hbar \omega}{2e}\right). \label{eq:tiengordonCP}
\end{equation}

\begin{figure}
	\centering
		\includegraphics[width = 1.00\columnwidth]{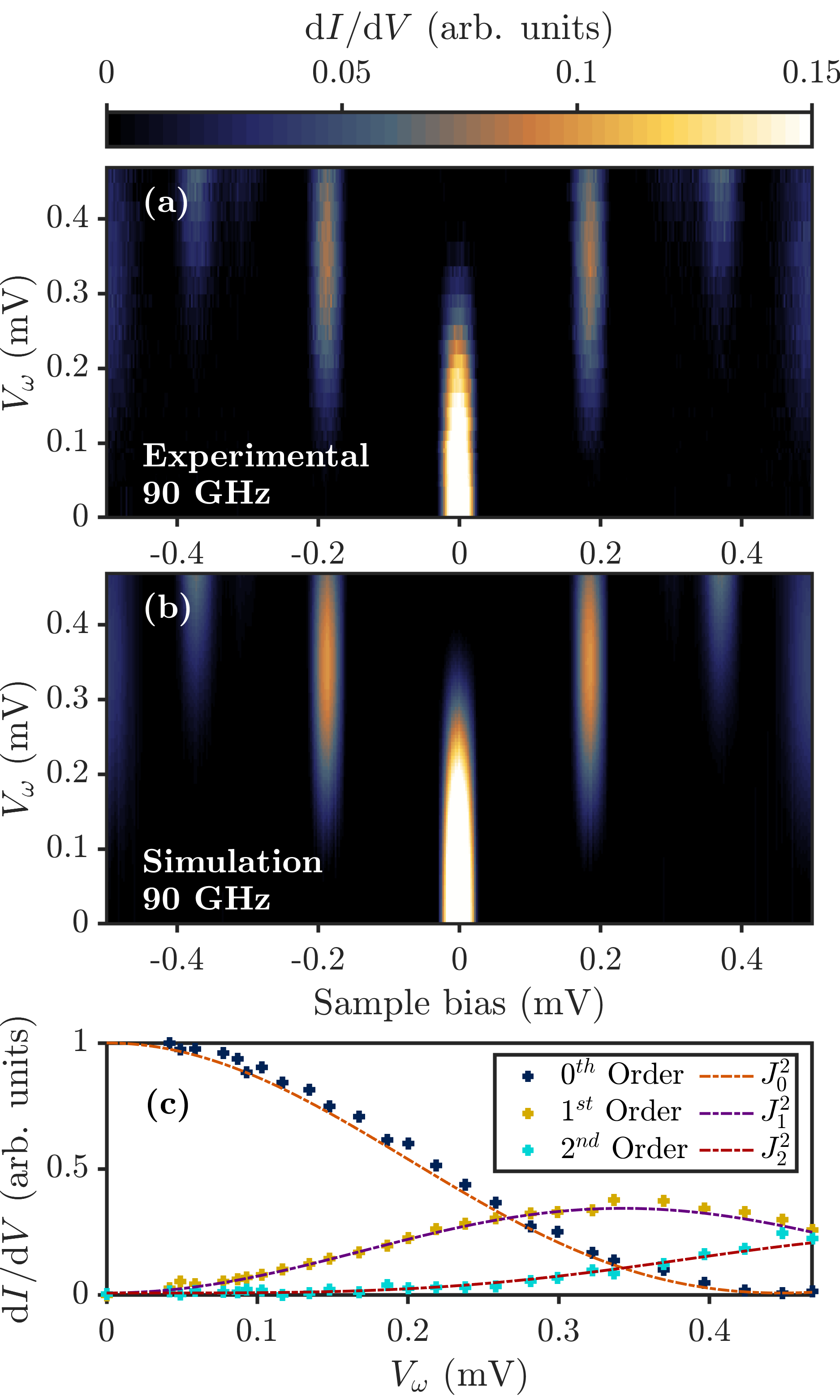}
	\caption{The Josephson effect under microwave irradiation: (a) Stacked contour plot of a series of conductance spectra of an SIS junction under irradiation by a 90\,GHz signal at varying $V_{\omega}$ with a focus on the Josephson effect. Satellite peaks now appear at an offset of $\hbar \omega / 2$. $G_N = 7.6\times10^{-3}\,G_0$. (b) Simulation of the data in panel a) using an experimental conductance spectrum and \eqref{eq:tiengordonCP}. (c) Constant $V_{0}$ profiles of the experimental data in panel a) at the principal Jospehson peak (dark blue) and the first (yellow) and second (cyan) satellites, nomalised to the maximum of the Josephson peak at $V_{\omega} = 0$. The dashed-dotted lines are the Bessel functions of zeroth (orange), first (purple), and second (red) order in $\frac{2 e V_{\omega}}{\hbar \omega}$.} 
	\label{fig:figure4}
\end{figure}

The only difference to \eqref{eq:tiengordon} is the replacement $e \to 2e$. Despite their stark difference in physical origin, the QP and the CP current show the same functional dependencies in their interactions with the microwave radiation. In another series of conductance spectra, shown as a contour plot in \figref{fig:figure4}, we measured the Josephson effect in the presence of a microwave signal of increasing amplitude. Experimental data is presented in \figref{fig:figure4}(a). The Josephson peak at zero bias is gradually fanning open with increasing $V_{\omega}$. Contributions from MARs and the coherence peaks soon begin to influence the relevant voltage range and restrict the measurement to small $V_{\omega}$. The observed pattern is a result of sequential CP tunneling in a low-capacitance junction and cannot be understood in terms of Shapiro steps, which require the phase difference across the junction to be a good quantum number\,\cite{si}. In the tunnel junction of an STM at ultra-low temperatures, the low junction capacitance leads to significant phase fluctuations, such that the Josephson effect has to be modelled as a sequential process within the $P(E)$-theory\,\cite{averin1990incoherent, devoret1990effect, ingold1994cooper, falci1991quasiparticle, grabert2015dynamical, roychowdhury2015microwave, ast2016sensing}. A simulation of the data using \eqref{eq:tiengordonCP} is shown in \figref{fig:figure4}(b). Also in this case, the peak profiles at constant $V_0$ follow the predicted Bessel function dependency with great accuracy, as can be seen in \figref{fig:figure4}(c). 

Given the series of features offset by integer fractions of $\hbar \omega$, it is tempting to think of the tunnel current as being carried by a series of dressed electron states with total charge $q = me$, where $m = 1$ for QP tunneling, $m = 2$ for CP tunneling and $m \geq 2$ for MARs, and each governed by an equation of the form of Eqs.\,(\ref{eq:tiengordon}) and (\ref{eq:tiengordonCP}) with the appropriate charge. Indeed, such an interpretation of the microwave radiation coupling to the total charge transferred between tip and sample during a tunnel process has been put forward\,\cite{roychowdhury2015microwave}. This idea is experimentally testable by considering MARs, which naturally contain tunnel processes transferring integer multiples of the elementary charge across the junction.

Tunneling in an SIS junction must be understood in the framework of MARs, the scattering processes permitting the conversion of a normal current into a supercurrent. An electron incident onto the superconductor from within the barrier is reflected as a hole, thereby transferring a charge of 2\textit{e} into the superconductor and forming a CP. As particles of opposite charge are traveling in opposite directions, MAR processes can result in the transfer of multiple elementary charges across the junction. There is thus no quasi-particle analogue to CPs in Andreev transport to which the microwave signal could be coupled and it is likely that the interactions of all particles participating in an Andreev process need to be considered explicitly to form a theoretical model of the process. This is the approach of the microwave-assisted MAR model in reference\,\cite{cuevas2002subharmonic}.

\begin{figure}
	\centering
		\includegraphics[width=1.0\columnwidth]{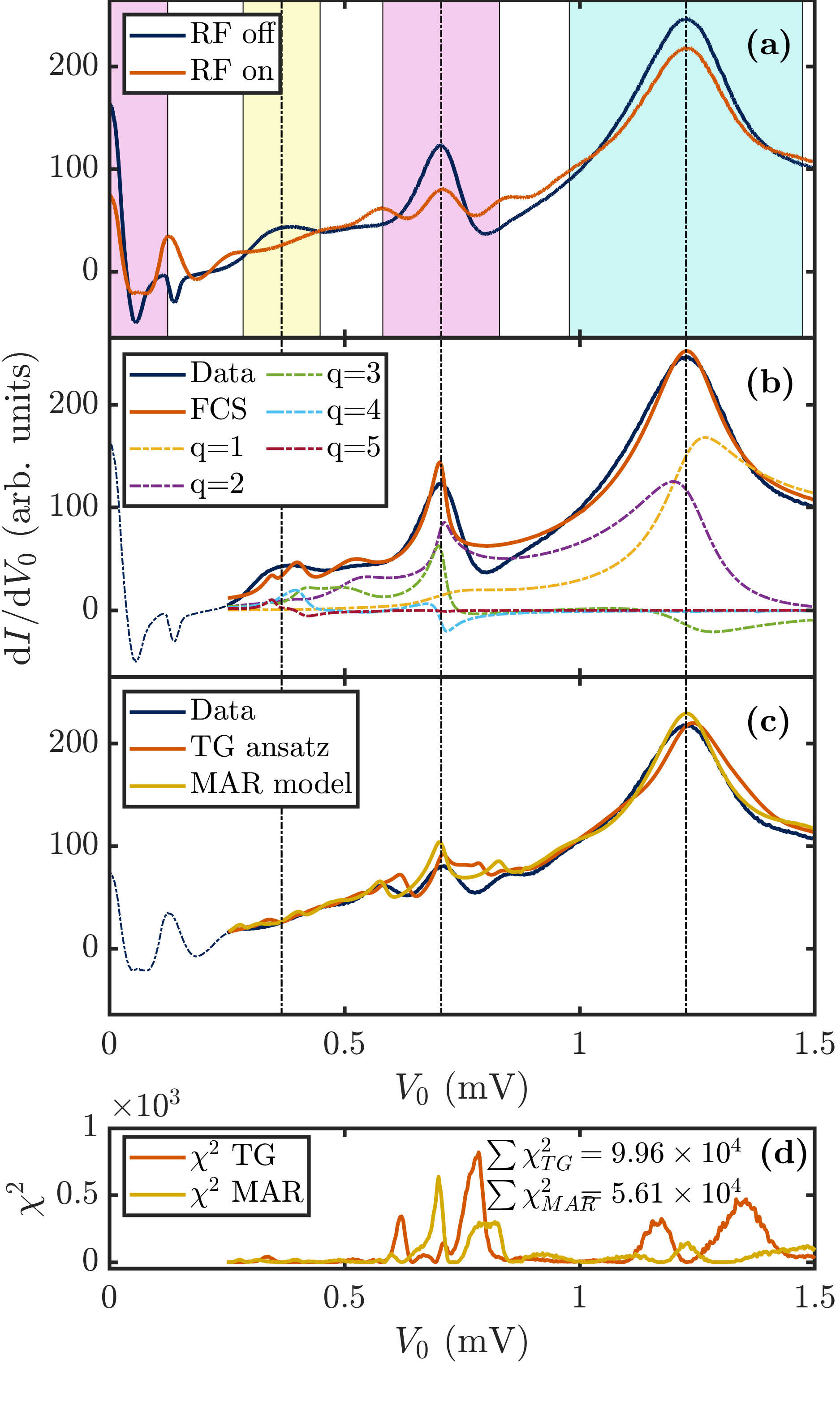}
	\caption{modeling the interaction with the microwave signal: (a) Conductance spectrum at $G_N = 1.07\, G_0$ in absence (dark blue) and presence (orange) of a microwave signal. The shaded areas mark energy offsets of $\hbar \omega$ (cyan), $\hbar \omega / 2$ (purple) and $\hbar \omega / 3$ (yellow). (b) Experimental conductance spectrum (dark blue) and FCS model thereof (orange). The individual contributions to the conductance from currents carrying \textit{q} = 1, 2, 3, 4, 5 charges are shown in yellow, purple, green, cyan, and red, respectively. The Josephson effect is not included in the model. (c) Comparison of the TG ansatz (orange) and exact calculation (yellow) to the experimental data (dark blue). (d) Square difference between the models and the experimental data. The MAR theory gives a significantly better description of the data than the TG Ansatz.}
	\label{fig:figure5}
\end{figure}

To investigate the effect of the microwave signal on Andreev transport, we increase the normal state conductance of the junction by reducing the tip-sample distance. A representative conductance spectrum of a high conductance junction with ($G_N = 1.07\,G_0$) can be found in \figref{fig:figure5}(a) in dark blue. In addition to the coherence peak located at $\approx$1.22\,mV, there are now a series of sub-gap features related to MARs. Most prominent are the first and second order MAR peaks at $\approx$0.7\,meV and $\approx$0.36\,meV, respectively. A spectrum of the same junction under microwave irradiation at 60\,GHz is shown in \figref{fig:figure5}(a) in orange. At first glance, the data seems to support the idea of a charge-sensitive measurement as replica from the first and second MAR are found at the correct offsets of $\approx\frac{\hbar \omega}{2 e}$ and $\approx\frac{\hbar \omega}{3 e}$, respectively.

A quantitative analysis of the data requires a thorough characterization of the junction. This means, ideally, to determine the number of conduction channels and their respective transmission probabilities, which are generically known as the junction PIN code. This can be done by analyzing the sub-gap structure in the absence of microwaves using the standard MAR theory and fitting procedures that are well described in the literature\,\cite{averin1995ac, cuevas1996hamiltonian, scheer1997conduction, scheer1998signature, cuevas1998microscopic, senkpiel2018single}. For the case of the data shown in \figref{fig:figure5}(a), the PIN code analysis finds a total of five transport channels with transmissions $\tau_i$ = [0.416, 0.293, 0.115, 0.114, 0.112] (see Supplemental Material for more details on the PIN code analysis)\,\cite{si}. This is consistent with the $d$-band nature of vanadium\,\cite{cuevas1998microscopic, scheer1998signature}. In addition to the junction PIN code, the amplitude of the microwave signal at the junction needs to be known. It is determined by acquiring a conductance spectrum in the tunneling regime, where \eqref{eq:tiengordon} can be applied, and performing a fit as described above.

Modeling the data according to the Tien-Gordon theory further requires knowledge of the magnitudes of current contributions carrying $m$ elementary charges. Such a decomposition of the total current is possible through full counting statistics (FCS), where it is found that multiple charge processes make a significant contribution to the total current across the junction and even dominate in the sub-gap regime at high conductance\,\cite{cuevas2003full, johansson2003full, cuevas2004dc}. The separate current contributions can then be treated using an adapted version of \eqref{eq:tiengordon} and the results compared to the MAR calculation\,\cite{cuevas2002subharmonic}.

The result of the FCS calculation as well as the individual contributions to the total conductance are shown in \figref{fig:figure5}(b). The discrepancy around the first MAR and additional structure below ca. 0.5\,mV bias can be attributed to the difficulty in correctly accounting for broadening effects at the ultra-low temperature of the experiment\,\cite{jack2016critical, ast2016sensing}. Applying \eqref{eq:tiengordon} to the FCS result, accounting for multiple charge transfers in the individual contributions, yields the orange curve in \figref{fig:figure5}(c) ('TG ansatz'). The total conductance obtained deviates strongly from the experimental data, shown in dark blue, even qualitatively. This approach is not able to correctly reproduce the location of the peaks in the experimental data, nor their magnitude. The microwave-assisted MAR model, shown in yellow in \figref{fig:figure5}(c), gives a much better description of the data. In particular, the location of the satellite peaks of the first MAR are predicted accurately. The error in their height can be traced by to the original issue overestimating the magnitude of the first MAR in the FCS model.

A robust measure of the performance of either model is found in the respective squared difference to the data, $\chi^2$. Plots of $\chi^2$ are shown in \figref{fig:figure5}(d) for both models. The poor performance of the TG ansatz around the first MAR is obvious. Significant shortcomings in the area of the coherence peak are also revealed. The aforementioned difficulties in modeling the peak broadening also lead to some discrepancies around the first Andreev reflection in the case of the MAR model, though smaller than for the TG ansatz. Across the entire range of the spectrum, the MAR model performs significantly better as can be gauged by summing the squared differences (see \figref{fig:figure5}(d)\,).

\section{Conclusions}

The essential difference between the Tien-Gordon and microwave-assisted MAR models lies in the interaction between DC transport and the microwave signal. The Tien-Gordon ansatz assumes a clear separation. This view of tunneling breaks down when higher order processes are included. The exact calculation, on the other hand, sees DC transport and microwave effects as linked at the most fundamental of levels. QPs interact with the microwave signal as part of the tunneling process, absorbing or emitting packets of energy. For the case of higher order processes such as MARs, a pair of initial and final states may be linked through several pathways involving different interactions with the microwave field. Transport in the presence of an AC driving signal must then be understood as a sum over histories of particles traveling back and forth across the barrier. The success of the MAR model can thus be seen as a direct consequence of multiple reflections occurring within the barrier and the interference of the particles involved.

An AC driving signal applied to a tunnel junction cannot be seen as only exciting fundamental modes within the sample, or as a simple broadening mechanism. The conductance of the junction will necessarily be modified. In the simplest cases, such as normal conducting samples and SIN or SIS junctions well below 0.1 $G_0$, this modification is well captured by the Tien-Gorden \eqref{eq:tiengordon} and is equivalent to a re-distribution of the density of states of the electrodes. This simple model breaks down when higher order effects become significant. In superconducting junctions in particular, this is the case as soon as the first Andreev reflection contributes significantly to the current. More elaborate models are needed to understand the junction behavior in these cases\,\cite{cuevas2002subharmonic}.

Microwave assisted tunneling is a significant expansion of the STM toolbox. By pulling back the curtain over higher order tunnel processes, a microwave signal may be used to reveal fine details about the nature of quantum mechanical transport which are inherently unavailable in a pure DC measurement. In the case of SIS junctions in particular, the interference between the microwave signal and the inherent AC components of the supercurrent may be a pathway to observing signatures of the AC Josephson effect in superconducting point contacts.

\begin{acknowledgments}
The authors would like to thank Alfredo Levy-Yeyati, Ciprian Padurariu, and Bj\"orn Kubala for insightful discussions. J.C.C. acknowledges funding from the Spanish Ministry of Economy and Competitiveness (MINECO) (contract No. FIS2017-84057-P). J.A. acknowledges support from the IQST and the German Science Foundation (DFG) under grant AN336/12-1. This work was funded in part by the ERC Consolidator Grant AbsoluteSpin (Grant No.\ 681164).
\end{acknowledgments}

\newpage

\begin{center}
	\textbf{Supplemental Material for: Microwave-assisted tunneling and interference effects in superconducting junctions under fast driving signals}
\end{center}

\renewcommand{\eqref}[1]{Eq.\,(\ref{#1})}

\makeatletter
	\renewcommand*{\@biblabel}[1]{[S#1]}
\makeatother
\renewcommand{\thefigure}{S\arabic{figure}}
\renewcommand{\theequation}{S\arabic{equation}}

\section*{Modeling the SIS tunnel conductance}

In the BCS formalism, the density of states of a superconducting electrode is given by

\begin{equation}
	\rho(\omega)  = \Re \left[ \frac{\omega + i\Gamma}{\sqrt{(\omega + i\Gamma)^2 - \Delta^2}} \right],
	\label{eq:bcs}
\end{equation}

\noindent where $\omega$ is the energy, $\Gamma$ a phenomenological broadening parameter, and $\Delta$ the order parameter. We model the SIS tunneling by convolving two densities of state of the form \eqref{eq:bcs}. The temperature is considered in a further convolution with a Fermi step function\,\cite{tinkham}.

\section*{Field-driven tunneling}

A rough estimation of the work performed by the microwave field in the experiment can be done on the basis of the cycle-averaged quiver energy of a free charge in a harmonic field. This quantity, also known as the ponderomotive energy, is given by:

\begin{equation}
	U_p = \frac{e^2 E_{\omega}^2}{4 m_e \omega^2},
	\label{eq:ponderomotive}
\end{equation}

\noindent where $e$ is the elementary charge, $E_{\omega}$ the electric field amplitude, $m_e$ the electron mass and $\omega$ the angular frequency of the driving signal. The other relevant energy scale in the problem is the photon energy $\hbar \omega$. As typical tip-sample distances in the STM range on the order of 1 nm and typical valued of $V_{\omega}$ reached on our experiment range on the order of 1 mV, we estimate $E_{\omega} \approx 1\times10^6$ V/m. Thus, at 90\,GHz

\begin{equation}
	U_p \approx 0.138\,\text{eV}\hspace{1cm}\hbar\omega \approx 8\times10^{-4}\,\text{eV}.
\end{equation}
	
The ponderomotive energy (\ref{eq:ponderomotive}) is thus nearly three orders of magnitude larger than the photon energy and the behavior of the junction is dominated by the action of the classical electrical field.

\section*{Dynamical Coulomb blockade \MakeLowercase{vs.} phase tunneling}

The Josephson-effect is generally described in terms of the global phase difference $\phi$ between the Ginzburg-Landau wave functions of the two electrodes\,\suppcite{tinkham}. The Josephson relation for the supercurrent

\begin{equation}
	I_J = I_C \sin(\Phi),
	\label{eq:josephsonRelationCurrent}
\end{equation}
\noindent where $I_J$ is the Josephson current and $I_C$ the critical current of the junction, and the phase evolution

\begin{equation}
	\partial_t \phi = 2\frac{e V}{\hbar},
	\label{eq:josephsonRelationPhase}
\end{equation}
	
\noindent where $V$ is the junction bias, predict that the Jospehson supercurrent will contain AC components at non-zero bias. When driven by a microwave signal, the total bias across the junction will be

\begin{equation}
	V(t) = V_0 + V_{\omega}\cos(\omega  t).
	\label{eq:voft}
\end{equation}

\noindent Integrating \eqref{eq:voft} according to \eqref{eq:josephsonRelationPhase} leads to the time-dependent phase difference

\begin{equation}
	\phi(t) = \phi_0 + \frac{2 e V_0}{\hbar}t + \frac{2 e V_{\omega}}{\hbar \omega}\sin(\omega t).
	\label{eq:rfPhaseEvolution}
\end{equation}

\noindent Substituting this result into the Josephson relation (\ref{eq:josephsonRelationCurrent}) yields the expression

\begin{equation}
	I_S = I_C \sum_n (-1)^n J_n \left( \frac{2 e V_{\omega}}{\hbar \omega}\right) \sin\left( \phi_0 + \frac{2 e V_0}{\hbar}t - n \omega t\right).
	\label{eq:shapiroSteps}
\end{equation}

Whenever the AC drive is synchronised with the phase rotation, the time dependent terms in the sine cancel, leading to a large DC contribution to the current, the famous Shapiro steps\,\suppcite{shapiro1963josephson}. The amplitude of the $n^{\text{th}}$ Shapiro step is

\begin{equation}
	\left| I_n \right| = I_0 \left| J_n\left( \frac{2 e V_{\omega}}{\hbar \omega}  \right) \right|.
	\label{eq:shapiroStepHeight}
\end{equation}

This line of reasoning relies on the existence of a well-defined phase difference between the two electrodes of the junction. This assumption does not necessarily hold for the STM: The low capacitances of the sharp probe tip leads to a significant charging energy. Combined with the cryogenic temperatures of most STM experiments, the junction enters the dynamical Coulomb blockade regime (DCB) which is characterised by large quantum fluctuations of the phase and the incoherent tunneling of Cooper pairs. The phase $\phi$ is thus no longer a good quantum number and a breakdown of \eqref{eq:shapiroSteps} might be expected.

A theoretical treatment of the Josephson effect in the DCB regime leads to the following relation for the Josephson current under AC driving\,\suppcite{falci1991quasiparticle}:

\begin{equation}
    I(V_{0}, V_{\omega}) = \sum^\infty_{n=-\infty} J_n^2 \left( \frac{2e\text{V}_{\omega}}{h \nu}\right ) I_0 \left( V_{0}-\frac{n h \nu}{2e}\right).
	\label{eq:tiengordonCP}
\end{equation}

\begin{figure}[!h]
	\centering
		\includegraphics[width=1.00\columnwidth]{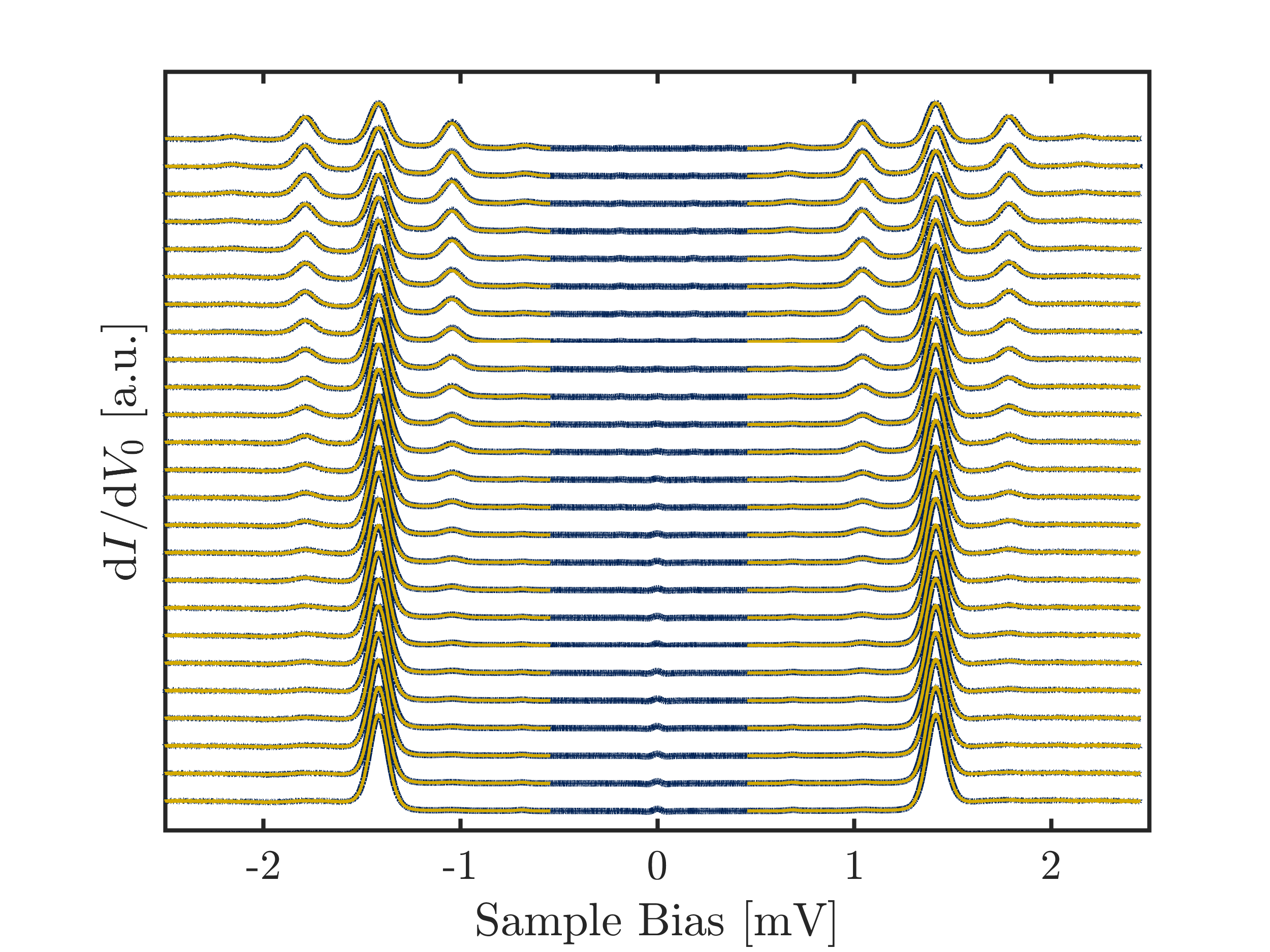}
		\caption{STM data acquired at increasingly high $V_{\omega}$ in dark blue with fits to the quasi-particle part of the spectrum according to the Tien-Gordon equation superimposed in yellow. Curves are offset for clarity. $G_N \approx 0.06 G_0$.}
	\label{fig:figureS1}
\end{figure}

The phenomenology of \eqref{eq:tiengordonCP} and \eqref{eq:shapiroSteps} is similar in the sense that both predict the emergence of satellite peaks to the Josephson current at the same offset of $\hbar \omega / 2$. The progression of the Josephson peak height and its replica, however, is vastly different in both cases and can serve as an excellent indicator that the experiment is indeed operating in the DCB regime.

To avoid any inherent bias towards one of the models, we extract $V_{\omega}$ independently by fitting the quasi-particle part of the spectrum using the Tien-Gordon equation (Eq. (1) of the main body text).  The resulting fits are shown in \figref{fig:figureS1}. 

\begin{figure}
	\centering
		\includegraphics[width=1.00\columnwidth]{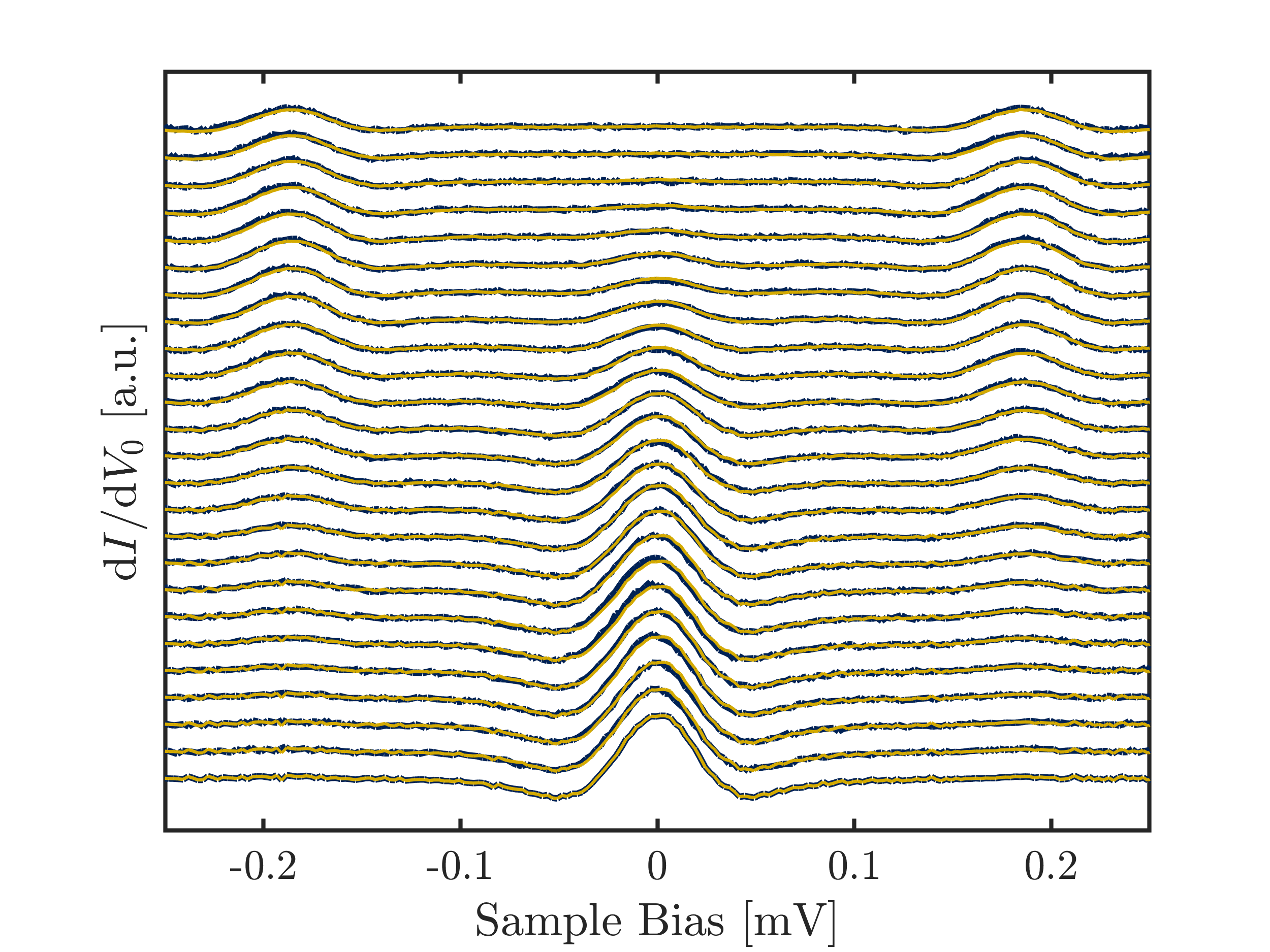}
		\caption{Detail from \figref{fig:figureS1} showing the Josephson effect at increasing values of $V_{\omega}$. Experimental data is shown in dark blue. Superimposed in yellow is the model from \eqref{eq:tiengordonCP} using only the amplitudes determined from the quasi-particle fits mentioned above and a spectrum at $V_{\omega} = 0$\,mV as input. Note that there are no free parameters in the modeling. Curves are offset for clarity.}
	\label{fig:figureS2}
\end{figure}

We can now attempt to model the Josephson effect using \eqref{eq:tiengordonCP} with only the independently determined values of $V_{\omega}$ and a reference spectrum at zero amplitude as inputs. The model thus contains no free parameters. The results are shown in \figref{fig:figureS2}. \eqref{eq:tiengordonCP} gives an excellent description of the experimental data. This serves as a first indicator that the experiment is indeed operating in the DCB regime.

\begin{figure}
	\centering
		\includegraphics[width=1.00\columnwidth]{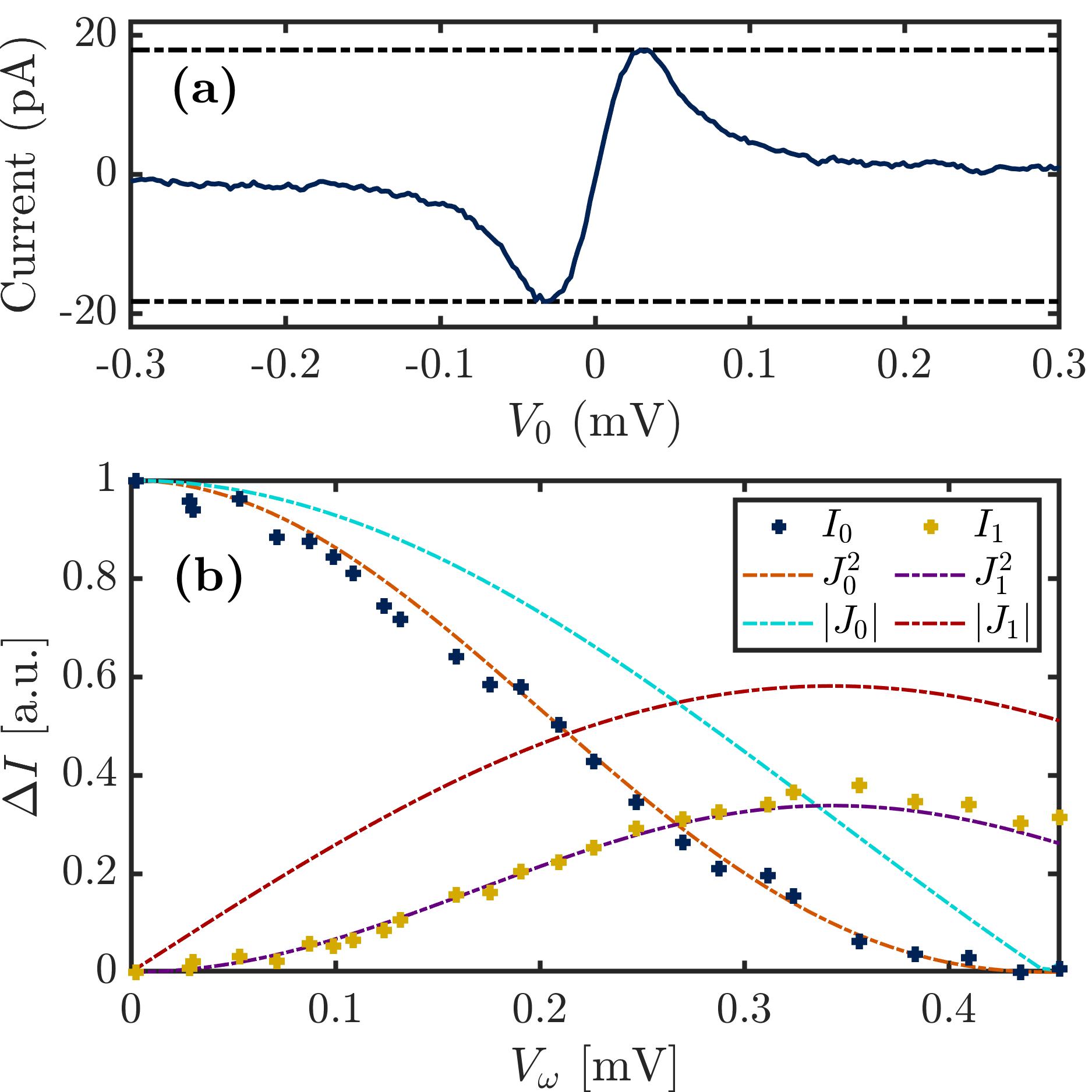}
	\caption{(a) The step height of the Josephson effect at zero bias and its first satellites were determined as shown by taking the difference between the minimum and maximum of the peak-and-dip feature in the current trace. (b) Step height of the Josephson effect and its first satellite as determined by the method described in panel a) with the value and square of the first two Bessel functions.}
	\label{fig:figureS3}
\end{figure}

Finally, we compare the step height of the Josephson feature in the current in dependence of $V_{\omega}$ to models in \eqref{eq:shapiroStepHeight} and \eqref{eq:tiengordonCP}. The step height is measured by taking the difference bewteen the miniumum and maximum of the peak-and-dip feature seen in the STM current for the Josephson effect and its first satellites at all values of $V_{\omega}$ (see \figref{fig:figureS3} (a) ). The experimental step height is shown in \ref{fig:figureS3}b) along with the predictions from models in \eqref{eq:shapiroStepHeight} and \eqref{eq:tiengordonCP}. It is clear that the observed satellite peaks are not consistent with Shapiro steps. The experiment is thus taking place in the DCB regime where $\phi$ is no longer a good quantum number.

\section*{PIN code determination}

The Multiple Andreev-reflection (MAR) model used in the analysis of the data presented in the main body manuscript is formulated for individual transport channels. However, a general point contact, such as encountered in the STM, is composed of several channels with varying transmissions. The MAR analysis can still be applied to our data when assuming that the channel transmissions are constant in the voltage window of interest and that transport channels themselves are independent of each other. The total current is then simply a sum over all transport channels.

The number $N$ of transport channels and their respective transmissions $\tau_j$, also known as the junction PIN code, is thus an important characteristic of any point contact. It can be determined from the sub-gap structure of MARs in a junction at high conductance. Higher order processes become increasingly important as the transmission increases and it can be shown that the prominence of the $n^{\text{th}}$ MAR scales with the power of $\tau_j^n$\,\suppcite{bratus1995theory}. The sub-gap structure is thus highly characteristic of the channel composition\,\suppcite{scheer1997conduction}.

\begin{figure}[!h]
	\centering
		\includegraphics[width=1.00\columnwidth]{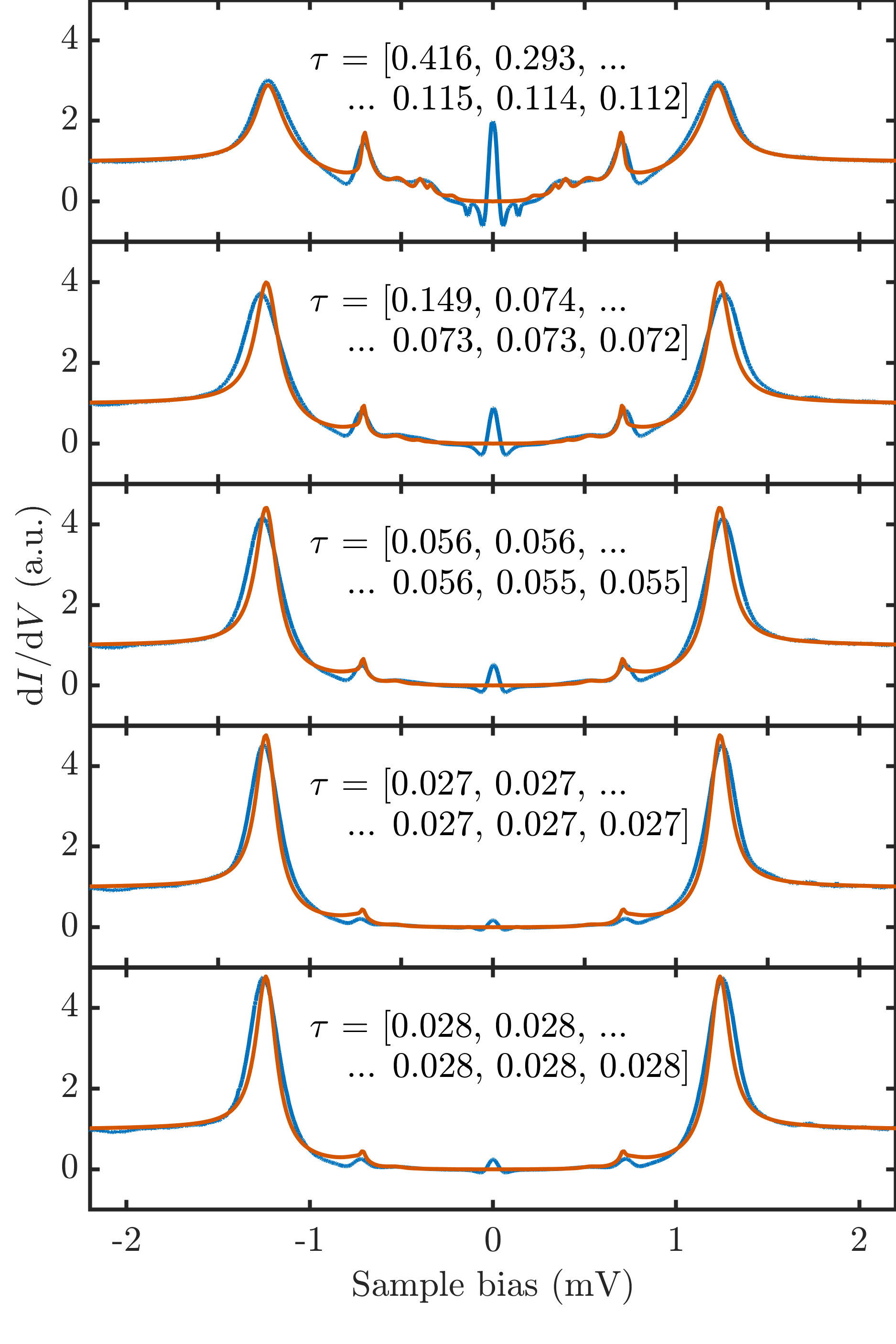}
	\caption{Experimental data (blue) and results from the PIN code search algorithm (orange) for several contacts with different channel compositions. Note that the Josephson effect is not included in the model and thus does not appear in the fit curve.}
	\label{fig:figureS4}
\end{figure}

We begin by extracting key parameters of the junction as described above. We acquire a spectrum at low-conductance and perform a fit using a convolution of two densities of states of the form of \eqref{eq:bcs} to determine $\Delta_{\text{Tip/Sample}}$ and $\Gamma_{\text{Tip/Sample}}$. A good description is reached with $\Delta_{\text{Sample}} = 710$\,$\mu e$V, $\Delta_{\text{Tip}} = 529$\,$\mu e$V, $\Gamma_{\text{Sample}} = 1\times10^{-5}$\,eV, and $\Gamma_{\text{Tip}} = 5.2\times10^{-5}$\,eV. These parameters are used to model the same junction at higher conductance. Taking care that the tip remains unchanged, we gradually increase the current setpoint and acquire a new spectrum with significant MAR contributions. We assume the total current in the junction to be a linear superposition of transport channels, such that\,\suppcite{cuevas1996hamiltonian, scheer1997conduction}

\begin{equation}
	I_{\text{Tot}}(V) = \sum_{j = 1}^N I_j(V, \tau_j).
	\label{eq:totalCurrent}
\end{equation}

For computational efficiency, we calculate a look-up table of current traces with varying transmissions $0 < \tau_j < 1$. We then model the total experimental current using \eqref{eq:totalCurrent} by summing entries in the look-up table. Owing to its partially filled $d$-shell, an atomic contact of V may sustain up to five transport channels\,\suppcite{cuevas1998microscopic, scheer1998signature}. A fitting algorithm attempts to replicate the experimental current trace using a sum of independent transport channels. To ensure the robustness of the final result, the algorithm performs a statistical search for a global minimum by repeatedly performing the same optimisation task beginning from many randomly selected starting values for the PIN code. A representative set of fitting results and their PIN code is shown in \figref{fig:figureS4}.


\begin{thebibliography}{37}%
\makeatletter
\providecommand \@ifxundefined [1]{%
 \@ifx{#1\undefined}
}%
\providecommand \@ifnum [1]{%
 \ifnum #1\expandafter \@firstoftwo
 \else \expandafter \@secondoftwo
 \fi
}%
\providecommand \@ifx [1]{%
 \ifx #1\expandafter \@firstoftwo
 \else \expandafter \@secondoftwo
 \fi
}%
\providecommand \natexlab [1]{#1}%
\providecommand \enquote  [1]{``#1''}%
\providecommand \bibnamefont  [1]{#1}%
\providecommand \bibfnamefont [1]{#1}%
\providecommand \citenamefont [1]{#1}%
\providecommand \href@noop [0]{\@secondoftwo}%
\providecommand \href [0]{\begingroup \@sanitize@url \@href}%
\providecommand \@href[1]{\@@startlink{#1}\@@href}%
\providecommand \@@href[1]{\endgroup#1\@@endlink}%
\providecommand \@sanitize@url [0]{\catcode `\\12\catcode `\$12\catcode
  `\&12\catcode `\#12\catcode `\^12\catcode `\_12\catcode `\%12\relax}%
\providecommand \@@startlink[1]{}%
\providecommand \@@endlink[0]{}%
\providecommand \url  [0]{\begingroup\@sanitize@url \@url }%
\providecommand \@url [1]{\endgroup\@href {#1}{\urlprefix }}%
\providecommand \urlprefix  [0]{URL }%
\providecommand \Eprint [0]{\href }%
\providecommand \doibase [0]{http://dx.doi.org/}%
\providecommand \selectlanguage [0]{\@gobble}%
\providecommand \bibinfo  [0]{\@secondoftwo}%
\providecommand \bibfield  [0]{\@secondoftwo}%
\providecommand \translation [1]{[#1]}%
\providecommand \BibitemOpen [0]{}%
\providecommand \bibitemStop [0]{}%
\providecommand \bibitemNoStop [0]{.\EOS\space}%
\providecommand \EOS [0]{\spacefactor3000\relax}%
\providecommand \BibitemShut  [1]{\csname bibitem#1\endcsname}%
\let\auto@bib@innerbib\@empty
\bibitem [{\citenamefont {Binnig}\ \emph {et~al.}(1982)\citenamefont {Binnig},
  \citenamefont {Rohrer}, \citenamefont {Gerber},\ and\ \citenamefont
  {Weibel}}]{binnig1982surface}%
  \BibitemOpen
  \bibfield  {author} {\bibinfo {author} {\bibfnamefont {G.}~\bibnamefont
  {Binnig}}, \bibinfo {author} {\bibfnamefont {H.}~\bibnamefont {Rohrer}},
  \bibinfo {author} {\bibfnamefont {C.}~\bibnamefont {Gerber}}, \ and\ \bibinfo
  {author} {\bibfnamefont {E.}~\bibnamefont {Weibel}},\ }\href@noop {}
  {\bibfield  {journal} {\bibinfo  {journal} {Phys. Rev. Lett.}\ }\textbf
  {\bibinfo {volume} {49}},\ \bibinfo {pages} {57} (\bibinfo {year}
  {1982})}\BibitemShut {NoStop}%
\bibitem [{\citenamefont {Loth}\ \emph {et~al.}(2010)\citenamefont {Loth},
  \citenamefont {Etzkorn}, \citenamefont {Lutz}, \citenamefont {Eigler},\ and\
  \citenamefont {Heinrich}}]{loth2010measurement}%
  \BibitemOpen
  \bibfield  {author} {\bibinfo {author} {\bibfnamefont {S.}~\bibnamefont
  {Loth}}, \bibinfo {author} {\bibfnamefont {M.}~\bibnamefont {Etzkorn}},
  \bibinfo {author} {\bibfnamefont {C.~P.}\ \bibnamefont {Lutz}}, \bibinfo
  {author} {\bibfnamefont {D.}~\bibnamefont {Eigler}}, \ and\ \bibinfo {author}
  {\bibfnamefont {A.~J.}\ \bibnamefont {Heinrich}},\ }\href@noop {} {\bibfield
  {journal} {\bibinfo  {journal} {Science}\ }\textbf {\bibinfo {volume}
  {329}},\ \bibinfo {pages} {1628} (\bibinfo {year} {2010})}\BibitemShut
  {NoStop}%
\bibitem [{\citenamefont {Terada}\ \emph {et~al.}(2010)\citenamefont {Terada},
  \citenamefont {Yoshida}, \citenamefont {Takeuchi},\ and\ \citenamefont
  {Shigekawa}}]{terada2010laser}%
  \BibitemOpen
  \bibfield  {author} {\bibinfo {author} {\bibfnamefont {Y.}~\bibnamefont
  {Terada}}, \bibinfo {author} {\bibfnamefont {S.}~\bibnamefont {Yoshida}},
  \bibinfo {author} {\bibfnamefont {O.}~\bibnamefont {Takeuchi}}, \ and\
  \bibinfo {author} {\bibfnamefont {H.}~\bibnamefont {Shigekawa}},\ }\href@noop
  {} {\bibfield  {journal} {\bibinfo  {journal} {Journal of Physics: Condensed
  Matter}\ }\textbf {\bibinfo {volume} {22}},\ \bibinfo {pages} {264008}
  (\bibinfo {year} {2010})}\BibitemShut {NoStop}%
\bibitem [{\citenamefont {Cocker}\ \emph {et~al.}(2013)\citenamefont {Cocker},
  \citenamefont {Jelic}, \citenamefont {Gupta}, \citenamefont {Molesky},
  \citenamefont {Burgess}, \citenamefont {De~Los~Reyes}, \citenamefont
  {Titova}, \citenamefont {Tsui}, \citenamefont {Freeman},\ and\ \citenamefont
  {Hegmann}}]{cocker2013ultrafast}%
  \BibitemOpen
  \bibfield  {author} {\bibinfo {author} {\bibfnamefont {T.~L.}\ \bibnamefont
  {Cocker}}, \bibinfo {author} {\bibfnamefont {V.}~\bibnamefont {Jelic}},
  \bibinfo {author} {\bibfnamefont {M.}~\bibnamefont {Gupta}}, \bibinfo
  {author} {\bibfnamefont {S.~J.}\ \bibnamefont {Molesky}}, \bibinfo {author}
  {\bibfnamefont {J.~A.~J.}\ \bibnamefont {Burgess}}, \bibinfo {author}
  {\bibfnamefont {G.}~\bibnamefont {De~Los~Reyes}}, \bibinfo {author}
  {\bibfnamefont {L.~V.}\ \bibnamefont {Titova}}, \bibinfo {author}
  {\bibfnamefont {Y.~Y.}\ \bibnamefont {Tsui}}, \bibinfo {author}
  {\bibfnamefont {M.~R.}\ \bibnamefont {Freeman}}, \ and\ \bibinfo {author}
  {\bibfnamefont {F.~A.}\ \bibnamefont {Hegmann}},\ }\href@noop {} {\bibfield
  {journal} {\bibinfo  {journal} {Nat. Photon.}\ }\textbf {\bibinfo {volume}
  {7}},\ \bibinfo {pages} {620} (\bibinfo {year} {2013})}\BibitemShut {NoStop}%
\bibitem [{\citenamefont {Yoshida}\ \emph {et~al.}(2014)\citenamefont
  {Yoshida}, \citenamefont {Aizawa}, \citenamefont {Wang}, \citenamefont
  {Oshima}, \citenamefont {Mera}, \citenamefont {Matsuyama}, \citenamefont
  {Oigawa}, \citenamefont {Takeuchi},\ and\ \citenamefont
  {Shigekawa}}]{yoshida2014probing}%
  \BibitemOpen
  \bibfield  {author} {\bibinfo {author} {\bibfnamefont {S.}~\bibnamefont
  {Yoshida}}, \bibinfo {author} {\bibfnamefont {Y.}~\bibnamefont {Aizawa}},
  \bibinfo {author} {\bibfnamefont {Z.-h.}\ \bibnamefont {Wang}}, \bibinfo
  {author} {\bibfnamefont {R.}~\bibnamefont {Oshima}}, \bibinfo {author}
  {\bibfnamefont {Y.}~\bibnamefont {Mera}}, \bibinfo {author} {\bibfnamefont
  {E.}~\bibnamefont {Matsuyama}}, \bibinfo {author} {\bibfnamefont
  {H.}~\bibnamefont {Oigawa}}, \bibinfo {author} {\bibfnamefont
  {O.}~\bibnamefont {Takeuchi}}, \ and\ \bibinfo {author} {\bibfnamefont
  {H.}~\bibnamefont {Shigekawa}},\ }\href@noop {} {\bibfield  {journal}
  {\bibinfo  {journal} {Nat. Nanotechnol.}\ }\textbf {\bibinfo {volume} {9}},\
  \bibinfo {pages} {588} (\bibinfo {year} {2014})}\BibitemShut {NoStop}%
\bibitem [{\citenamefont {Rashidi}\ \emph {et~al.}(2016)\citenamefont
  {Rashidi}, \citenamefont {Burgess}, \citenamefont {Taucer}, \citenamefont
  {Achal}, \citenamefont {Pitters}, \citenamefont {Loth},\ and\ \citenamefont
  {Wolkow}}]{rashidi2016time}%
  \BibitemOpen
  \bibfield  {author} {\bibinfo {author} {\bibfnamefont {M.}~\bibnamefont
  {Rashidi}}, \bibinfo {author} {\bibfnamefont {J.~A.}\ \bibnamefont
  {Burgess}}, \bibinfo {author} {\bibfnamefont {M.}~\bibnamefont {Taucer}},
  \bibinfo {author} {\bibfnamefont {R.}~\bibnamefont {Achal}}, \bibinfo
  {author} {\bibfnamefont {J.~L.}\ \bibnamefont {Pitters}}, \bibinfo {author}
  {\bibfnamefont {S.}~\bibnamefont {Loth}}, \ and\ \bibinfo {author}
  {\bibfnamefont {R.~A.}\ \bibnamefont {Wolkow}},\ }\href@noop {} {\bibfield
  {journal} {\bibinfo  {journal} {Nat. Commun.}\ }\textbf {\bibinfo {volume}
  {7}},\ \bibinfo {pages} {13258} (\bibinfo {year} {2016})}\BibitemShut
  {NoStop}%
\bibitem [{\citenamefont {Cocker}\ \emph {et~al.}(2016)\citenamefont {Cocker},
  \citenamefont {Peller}, \citenamefont {Yu}, \citenamefont {Repp},\ and\
  \citenamefont {Huber}}]{cocker2016tracking}%
  \BibitemOpen
  \bibfield  {author} {\bibinfo {author} {\bibfnamefont {T.~L.}\ \bibnamefont
  {Cocker}}, \bibinfo {author} {\bibfnamefont {D.}~\bibnamefont {Peller}},
  \bibinfo {author} {\bibfnamefont {P.}~\bibnamefont {Yu}}, \bibinfo {author}
  {\bibfnamefont {J.}~\bibnamefont {Repp}}, \ and\ \bibinfo {author}
  {\bibfnamefont {R.}~\bibnamefont {Huber}},\ }\href@noop {} {\bibfield
  {journal} {\bibinfo  {journal} {Nature}\ }\textbf {\bibinfo {volume} {539}},\
  \bibinfo {pages} {263} (\bibinfo {year} {2016})}\BibitemShut {NoStop}%
\bibitem [{\citenamefont {Garg}\ and\ \citenamefont
  {Kern}(2019)}]{garg2019attosecond}%
  \BibitemOpen
  \bibfield  {author} {\bibinfo {author} {\bibfnamefont {M.}~\bibnamefont
  {Garg}}\ and\ \bibinfo {author} {\bibfnamefont {K.}~\bibnamefont {Kern}},\
  }\href@noop {} {\bibfield  {journal} {\bibinfo  {journal} {Science}\ ,\
  \bibinfo {pages} {eaaz1098}} (\bibinfo {year} {2019})}\BibitemShut {NoStop}%
\bibitem [{\citenamefont {Shapiro}(1963)}]{shapiro1963josephson}%
  \BibitemOpen
  \bibfield  {author} {\bibinfo {author} {\bibfnamefont {S.}~\bibnamefont
  {Shapiro}},\ }\href@noop {} {\bibfield  {journal} {\bibinfo  {journal} {Phys.
  Rev. Lett.}\ }\textbf {\bibinfo {volume} {11}},\ \bibinfo {pages} {80}
  (\bibinfo {year} {1963})}\BibitemShut {NoStop}%
\bibitem [{\citenamefont {Tien}\ and\ \citenamefont
  {Gordon}(1963)}]{tien1963multiphoton}%
  \BibitemOpen
  \bibfield  {author} {\bibinfo {author} {\bibfnamefont {P.}~\bibnamefont
  {Tien}}\ and\ \bibinfo {author} {\bibfnamefont {J.}~\bibnamefont {Gordon}},\
  }\href@noop {} {\bibfield  {journal} {\bibinfo  {journal} {Phys. Rev.}\
  }\textbf {\bibinfo {volume} {129}},\ \bibinfo {pages} {647} (\bibinfo {year}
  {1963})}\BibitemShut {NoStop}%
\bibitem [{\citenamefont {Falci}\ \emph {et~al.}(1991)\citenamefont {Falci},
  \citenamefont {Bubanja},\ and\ \citenamefont
  {Sch{\"o}n}}]{falci1991quasiparticle}%
  \BibitemOpen
  \bibfield  {author} {\bibinfo {author} {\bibfnamefont {G.}~\bibnamefont
  {Falci}}, \bibinfo {author} {\bibfnamefont {V.}~\bibnamefont {Bubanja}}, \
  and\ \bibinfo {author} {\bibfnamefont {G.}~\bibnamefont {Sch{\"o}n}},\
  }\href@noop {} {\bibfield  {journal} {\bibinfo  {journal} {Z. Phys. B}\
  }\textbf {\bibinfo {volume} {85}},\ \bibinfo {pages} {451} (\bibinfo {year}
  {1991})}\BibitemShut {NoStop}%
\bibitem [{\citenamefont {Chauvin}\ \emph {et~al.}(2006)\citenamefont
  {Chauvin}, \citenamefont {Vom~Stein}, \citenamefont {Pothier}, \citenamefont
  {Joyez}, \citenamefont {Huber}, \citenamefont {Esteve},\ and\ \citenamefont
  {Urbina}}]{chauvin2006superconducting}%
  \BibitemOpen
  \bibfield  {author} {\bibinfo {author} {\bibfnamefont {M.}~\bibnamefont
  {Chauvin}}, \bibinfo {author} {\bibfnamefont {P.}~\bibnamefont {Vom~Stein}},
  \bibinfo {author} {\bibfnamefont {H.}~\bibnamefont {Pothier}}, \bibinfo
  {author} {\bibfnamefont {P.}~\bibnamefont {Joyez}}, \bibinfo {author}
  {\bibfnamefont {M.}~\bibnamefont {Huber}}, \bibinfo {author} {\bibfnamefont
  {D.}~\bibnamefont {Esteve}}, \ and\ \bibinfo {author} {\bibfnamefont
  {C.}~\bibnamefont {Urbina}},\ }\href@noop {} {\bibfield  {journal} {\bibinfo
  {journal} {Phys. Rev. Lett.}\ }\textbf {\bibinfo {volume} {97}},\ \bibinfo
  {pages} {067006} (\bibinfo {year} {2006})}\BibitemShut {NoStop}%
\bibitem [{\citenamefont {Roychowdhury}\ \emph {et~al.}(2015)\citenamefont
  {Roychowdhury}, \citenamefont {Dreyer}, \citenamefont {Anderson},
  \citenamefont {Lobb},\ and\ \citenamefont
  {Wellstood}}]{roychowdhury2015microwave}%
  \BibitemOpen
  \bibfield  {author} {\bibinfo {author} {\bibfnamefont {A.}~\bibnamefont
  {Roychowdhury}}, \bibinfo {author} {\bibfnamefont {M.}~\bibnamefont
  {Dreyer}}, \bibinfo {author} {\bibfnamefont {J.}~\bibnamefont {Anderson}},
  \bibinfo {author} {\bibfnamefont {C.}~\bibnamefont {Lobb}}, \ and\ \bibinfo
  {author} {\bibfnamefont {F.}~\bibnamefont {Wellstood}},\ }\href@noop {}
  {\bibfield  {journal} {\bibinfo  {journal} {Phys. Rev. Applied}\ }\textbf
  {\bibinfo {volume} {4}},\ \bibinfo {pages} {034011} (\bibinfo {year}
  {2015})}\BibitemShut {NoStop}%
\bibitem [{\citenamefont {Cuevas}\ \emph {et~al.}(2002)\citenamefont {Cuevas},
  \citenamefont {Heurich}, \citenamefont {Mart{\'\i}n-Rodero}, \citenamefont
  {Levy~Yeyati},\ and\ \citenamefont {Sch{\"o}n}}]{cuevas2002subharmonic}%
  \BibitemOpen
  \bibfield  {author} {\bibinfo {author} {\bibfnamefont {J.~C.}\ \bibnamefont
  {Cuevas}}, \bibinfo {author} {\bibfnamefont {J.}~\bibnamefont {Heurich}},
  \bibinfo {author} {\bibfnamefont {A.}~\bibnamefont {Mart{\'\i}n-Rodero}},
  \bibinfo {author} {\bibfnamefont {A.}~\bibnamefont {Levy~Yeyati}}, \ and\
  \bibinfo {author} {\bibfnamefont {G.}~\bibnamefont {Sch{\"o}n}},\ }\href@noop
  {} {\bibfield  {journal} {\bibinfo  {journal} {Phys. Rev. Lett.}\ }\textbf
  {\bibinfo {volume} {88}},\ \bibinfo {pages} {157001} (\bibinfo {year}
  {2002})}\BibitemShut {NoStop}%
\bibitem [{\citenamefont {Davies}\ and\ \citenamefont
  {Lambert}(1980)}]{davies1980surface}%
  \BibitemOpen
  \bibfield  {author} {\bibinfo {author} {\bibfnamefont {P.~W.}\ \bibnamefont
  {Davies}}\ and\ \bibinfo {author} {\bibfnamefont {R.~M.}\ \bibnamefont
  {Lambert}},\ }\href@noop {} {\bibfield  {journal} {\bibinfo  {journal} {Surf.
  Sci.}\ }\textbf {\bibinfo {volume} {95}},\ \bibinfo {pages} {571} (\bibinfo
  {year} {1980})}\BibitemShut {NoStop}%
\bibitem [{\citenamefont {Foord}\ \emph {et~al.}(1983)\citenamefont {Foord},
  \citenamefont {Reed},\ and\ \citenamefont {Lambert}}]{foord1983the}%
  \BibitemOpen
  \bibfield  {author} {\bibinfo {author} {\bibfnamefont {J.~S.}\ \bibnamefont
  {Foord}}, \bibinfo {author} {\bibfnamefont {A.~P.~C.}\ \bibnamefont {Reed}},
  \ and\ \bibinfo {author} {\bibfnamefont {R.~M.}\ \bibnamefont {Lambert}},\
  }\href@noop {} {\bibfield  {journal} {\bibinfo  {journal} {Surf. Sci.}\
  }\textbf {\bibinfo {volume} {129}},\ \bibinfo {pages} {79} (\bibinfo {year}
  {1983})}\BibitemShut {NoStop}%
\bibitem [{si()}]{si}%
  \BibitemOpen
  \href@noop {} {}\bibinfo {note} {See Supplemental Material for further
  details on the modeling of the tunneling conductance, a discussion of
  field-driven tunneling, an analysis of the dynamical Coulomb blockade vs.
  Shapiro steps, and a description of the determination of the junction PIN
  code.}\BibitemShut {Stop}%
\bibitem [{\citenamefont {Soerensen}\ \emph {et~al.}(1974)\citenamefont
  {Soerensen}, \citenamefont {Kofoed}, \citenamefont {Pedersen},\ and\
  \citenamefont {Shapiro}}]{soerensen1974microwave}%
  \BibitemOpen
  \bibfield  {author} {\bibinfo {author} {\bibfnamefont {O.~H.}\ \bibnamefont
  {Soerensen}}, \bibinfo {author} {\bibfnamefont {B.}~\bibnamefont {Kofoed}},
  \bibinfo {author} {\bibfnamefont {N.~F.}\ \bibnamefont {Pedersen}}, \ and\
  \bibinfo {author} {\bibfnamefont {S.}~\bibnamefont {Shapiro}},\ }\href@noop
  {} {\bibfield  {journal} {\bibinfo  {journal} {Phys. Rev. B}\ }\textbf
  {\bibinfo {volume} {9}},\ \bibinfo {pages} {3746} (\bibinfo {year}
  {1974})}\BibitemShut {NoStop}%
\bibitem [{\citenamefont {Kouwenhoven}\ \emph {et~al.}(1994)\citenamefont
  {Kouwenhoven}, \citenamefont {Jauhar}, \citenamefont {McCormick},
  \citenamefont {Dixon}, \citenamefont {McEuen}, \citenamefont {Nazarov},
  \citenamefont {van~der Vaart},\ and\ \citenamefont
  {Foxon}}]{kouwenhoven1994photon}%
  \BibitemOpen
  \bibfield  {author} {\bibinfo {author} {\bibfnamefont {L.~P.}\ \bibnamefont
  {Kouwenhoven}}, \bibinfo {author} {\bibfnamefont {S.}~\bibnamefont {Jauhar}},
  \bibinfo {author} {\bibfnamefont {K.}~\bibnamefont {McCormick}}, \bibinfo
  {author} {\bibfnamefont {D.}~\bibnamefont {Dixon}}, \bibinfo {author}
  {\bibfnamefont {P.~L.}\ \bibnamefont {McEuen}}, \bibinfo {author}
  {\bibfnamefont {Y.~V.}\ \bibnamefont {Nazarov}}, \bibinfo {author}
  {\bibfnamefont {N.~C.}\ \bibnamefont {van~der Vaart}}, \ and\ \bibinfo
  {author} {\bibfnamefont {C.~T.}\ \bibnamefont {Foxon}},\ }\href@noop {}
  {\bibfield  {journal} {\bibinfo  {journal} {Phys. Rev. B}\ }\textbf {\bibinfo
  {volume} {50}},\ \bibinfo {pages} {2019} (\bibinfo {year}
  {1994})}\BibitemShut {NoStop}%
\bibitem [{\citenamefont {de~Graaf}\ \emph {et~al.}(2013)\citenamefont
  {de~Graaf}, \citenamefont {Lepp\"akangas}, \citenamefont {Adamyan},
  \citenamefont {Danilov}, \citenamefont {Lindstr\"om}, \citenamefont
  {Fogelstr\"om}, \citenamefont {Bauch}, \citenamefont {Johansson},\ and\
  \citenamefont {Kubatkin}}]{degraaf2013charge}%
  \BibitemOpen
  \bibfield  {author} {\bibinfo {author} {\bibfnamefont {S.~E.}\ \bibnamefont
  {de~Graaf}}, \bibinfo {author} {\bibfnamefont {J.}~\bibnamefont
  {Lepp\"akangas}}, \bibinfo {author} {\bibfnamefont {A.}~\bibnamefont
  {Adamyan}}, \bibinfo {author} {\bibfnamefont {A.~V.}\ \bibnamefont
  {Danilov}}, \bibinfo {author} {\bibfnamefont {T.}~\bibnamefont
  {Lindstr\"om}}, \bibinfo {author} {\bibfnamefont {M.}~\bibnamefont
  {Fogelstr\"om}}, \bibinfo {author} {\bibfnamefont {T.}~\bibnamefont {Bauch}},
  \bibinfo {author} {\bibfnamefont {G.}~\bibnamefont {Johansson}}, \ and\
  \bibinfo {author} {\bibfnamefont {S.~E.}\ \bibnamefont {Kubatkin}},\
  }\href@noop {} {\bibfield  {journal} {\bibinfo  {journal} {Phys. Rev. Lett.}\
  }\textbf {\bibinfo {volume} {111}},\ \bibinfo {pages} {137002} (\bibinfo
  {year} {2013})}\BibitemShut {NoStop}%
\bibitem [{\citenamefont {Kampfrath}\ \emph {et~al.}(2013)\citenamefont
  {Kampfrath}, \citenamefont {Tanaka},\ and\ \citenamefont
  {Nelson}}]{kampfrath2013resonant}%
  \BibitemOpen
  \bibfield  {author} {\bibinfo {author} {\bibfnamefont {T.}~\bibnamefont
  {Kampfrath}}, \bibinfo {author} {\bibfnamefont {K.}~\bibnamefont {Tanaka}}, \
  and\ \bibinfo {author} {\bibfnamefont {K.~A.}\ \bibnamefont {Nelson}},\
  }\href@noop {} {\bibfield  {journal} {\bibinfo  {journal} {Nat. Photon.}\
  }\textbf {\bibinfo {volume} {7}},\ \bibinfo {pages} {680} (\bibinfo {year}
  {2013})}\BibitemShut {NoStop}%
\bibitem [{\citenamefont {Tinkham}(1996)}]{tinkham}%
  \BibitemOpen
  \bibfield  {author} {\bibinfo {author} {\bibfnamefont {M.}~\bibnamefont
  {Tinkham}},\ }\href@noop {} {\emph {\bibinfo {title} {Introduction to
  Superconductivity}}},\ \bibinfo {edition} {2nd}\ ed.\ (\bibinfo  {publisher}
  {Dover Publications Inc.},\ \bibinfo {year} {1996})\BibitemShut {NoStop}%
\bibitem [{\citenamefont {Grabert}(2015)}]{grabert2015dynamical}%
  \BibitemOpen
  \bibfield  {author} {\bibinfo {author} {\bibfnamefont {H.}~\bibnamefont
  {Grabert}},\ }\href@noop {} {\bibfield  {journal} {\bibinfo  {journal} {Phys.
  Rev. B}\ }\textbf {\bibinfo {volume} {92}},\ \bibinfo {pages} {245433}
  (\bibinfo {year} {2015})}\BibitemShut {NoStop}%
\bibitem [{\citenamefont {Averin}\ \emph {et~al.}(1990)\citenamefont {Averin},
  \citenamefont {Nazarov},\ and\ \citenamefont
  {Odintsov}}]{averin1990incoherent}%
  \BibitemOpen
  \bibfield  {author} {\bibinfo {author} {\bibfnamefont {D.~V.}\ \bibnamefont
  {Averin}}, \bibinfo {author} {\bibfnamefont {Y.}~\bibnamefont {Nazarov}}, \
  and\ \bibinfo {author} {\bibfnamefont {A.~A.}\ \bibnamefont {Odintsov}},\
  }\href {\doibase http://dx.doi.org/10.1016/S0921-4526(09)80058-6} {\bibfield
  {journal} {\bibinfo  {journal} {Physica B}\ }\textbf {\bibinfo {volume}
  {165-166}},\ \bibinfo {pages} {945} (\bibinfo {year} {1990})}\BibitemShut
  {NoStop}%
\bibitem [{\citenamefont {Devoret}\ \emph {et~al.}(1990)\citenamefont
  {Devoret}, \citenamefont {Esteve}, \citenamefont {Grabert}, \citenamefont
  {Ingold}, \citenamefont {Pothier},\ and\ \citenamefont
  {Urbina}}]{devoret1990effect}%
  \BibitemOpen
  \bibfield  {author} {\bibinfo {author} {\bibfnamefont {M.~H.}\ \bibnamefont
  {Devoret}}, \bibinfo {author} {\bibfnamefont {D.}~\bibnamefont {Esteve}},
  \bibinfo {author} {\bibfnamefont {H.}~\bibnamefont {Grabert}}, \bibinfo
  {author} {\bibfnamefont {G.}~\bibnamefont {Ingold}}, \bibinfo {author}
  {\bibfnamefont {H.}~\bibnamefont {Pothier}}, \ and\ \bibinfo {author}
  {\bibfnamefont {C.}~\bibnamefont {Urbina}},\ }\href@noop {} {\bibfield
  {journal} {\bibinfo  {journal} {Phys. Rev. Lett.}\ }\textbf {\bibinfo
  {volume} {64}},\ \bibinfo {pages} {1824} (\bibinfo {year}
  {1990})}\BibitemShut {NoStop}%
\bibitem [{\citenamefont {Ingold}\ \emph {et~al.}(1994)\citenamefont {Ingold},
  \citenamefont {Grabert},\ and\ \citenamefont {Eberhardt}}]{ingold1994cooper}%
  \BibitemOpen
  \bibfield  {author} {\bibinfo {author} {\bibfnamefont {G.}~\bibnamefont
  {Ingold}}, \bibinfo {author} {\bibfnamefont {H.}~\bibnamefont {Grabert}}, \
  and\ \bibinfo {author} {\bibfnamefont {U.}~\bibnamefont {Eberhardt}},\
  }\href@noop {} {\bibfield  {journal} {\bibinfo  {journal} {Phys. Rev. B}\
  }\textbf {\bibinfo {volume} {50}},\ \bibinfo {pages} {395} (\bibinfo {year}
  {1994})}\BibitemShut {NoStop}%
\bibitem [{\citenamefont {Ast}\ \emph {et~al.}(2016)\citenamefont {Ast},
  \citenamefont {J{\"a}ck}, \citenamefont {Senkpiel}, \citenamefont {Eltschka},
  \citenamefont {Etzkorn}, \citenamefont {Ankerhold},\ and\ \citenamefont
  {Kern}}]{ast2016sensing}%
  \BibitemOpen
  \bibfield  {author} {\bibinfo {author} {\bibfnamefont {C.~R.}\ \bibnamefont
  {Ast}}, \bibinfo {author} {\bibfnamefont {B.}~\bibnamefont {J{\"a}ck}},
  \bibinfo {author} {\bibfnamefont {J.}~\bibnamefont {Senkpiel}}, \bibinfo
  {author} {\bibfnamefont {M.}~\bibnamefont {Eltschka}}, \bibinfo {author}
  {\bibfnamefont {M.}~\bibnamefont {Etzkorn}}, \bibinfo {author} {\bibfnamefont
  {J.}~\bibnamefont {Ankerhold}}, \ and\ \bibinfo {author} {\bibfnamefont
  {K.}~\bibnamefont {Kern}},\ }\href@noop {} {\bibfield  {journal} {\bibinfo
  {journal} {Nat. Commun.}\ }\textbf {\bibinfo {volume} {7}},\ \bibinfo {pages}
  {13009} (\bibinfo {year} {2016})}\BibitemShut {NoStop}%
\bibitem [{\citenamefont {Averin}\ and\ \citenamefont
  {Bardas}(1995)}]{averin1995ac}%
  \BibitemOpen
  \bibfield  {author} {\bibinfo {author} {\bibfnamefont {D.}~\bibnamefont
  {Averin}}\ and\ \bibinfo {author} {\bibfnamefont {A.}~\bibnamefont
  {Bardas}},\ }\href@noop {} {\bibfield  {journal} {\bibinfo  {journal} {Phys.
  Rev. Lett.}\ }\textbf {\bibinfo {volume} {75}},\ \bibinfo {pages} {1831}
  (\bibinfo {year} {1995})}\BibitemShut {NoStop}%
\bibitem [{\citenamefont {Cuevas}\ \emph {et~al.}(1996)\citenamefont {Cuevas},
  \citenamefont {Mart{\'\i}n-Rodero},\ and\ \citenamefont
  {Levy~Yeyati}}]{cuevas1996hamiltonian}%
  \BibitemOpen
  \bibfield  {author} {\bibinfo {author} {\bibfnamefont {J.~C.}\ \bibnamefont
  {Cuevas}}, \bibinfo {author} {\bibfnamefont {A.}~\bibnamefont
  {Mart{\'\i}n-Rodero}}, \ and\ \bibinfo {author} {\bibfnamefont
  {A.}~\bibnamefont {Levy~Yeyati}},\ }\href@noop {} {\bibfield  {journal}
  {\bibinfo  {journal} {Phys. Rev. B}\ }\textbf {\bibinfo {volume} {54}},\
  \bibinfo {pages} {7366} (\bibinfo {year} {1996})}\BibitemShut {NoStop}%
\bibitem [{\citenamefont {Scheer}\ \emph {et~al.}(1997)\citenamefont {Scheer},
  \citenamefont {Joyez}, \citenamefont {Esteve}, \citenamefont {Urbina},\ and\
  \citenamefont {Devoret}}]{scheer1997conduction}%
  \BibitemOpen
  \bibfield  {author} {\bibinfo {author} {\bibfnamefont {E.}~\bibnamefont
  {Scheer}}, \bibinfo {author} {\bibfnamefont {P.}~\bibnamefont {Joyez}},
  \bibinfo {author} {\bibfnamefont {D.}~\bibnamefont {Esteve}}, \bibinfo
  {author} {\bibfnamefont {C.}~\bibnamefont {Urbina}}, \ and\ \bibinfo {author}
  {\bibfnamefont {M.~H.}\ \bibnamefont {Devoret}},\ }\href@noop {} {\bibfield
  {journal} {\bibinfo  {journal} {Phys. Rev. Lett.}\ }\textbf {\bibinfo
  {volume} {78}},\ \bibinfo {pages} {3535} (\bibinfo {year}
  {1997})}\BibitemShut {NoStop}%
\bibitem [{\citenamefont {Scheer}\ \emph {et~al.}(1998)\citenamefont {Scheer},
  \citenamefont {Agra{\"\i}t}, \citenamefont {Cuevas}, \citenamefont
  {Levy~Yeyati}, \citenamefont {Ludoph}, \citenamefont {Mart{\'\i}n-Rodero},
  \citenamefont {Bollinger}, \citenamefont {van Ruitenbeek},\ and\
  \citenamefont {Urbina}}]{scheer1998signature}%
  \BibitemOpen
  \bibfield  {author} {\bibinfo {author} {\bibfnamefont {E.}~\bibnamefont
  {Scheer}}, \bibinfo {author} {\bibfnamefont {N.}~\bibnamefont {Agra{\"\i}t}},
  \bibinfo {author} {\bibfnamefont {J.~C.}\ \bibnamefont {Cuevas}}, \bibinfo
  {author} {\bibfnamefont {A.}~\bibnamefont {Levy~Yeyati}}, \bibinfo {author}
  {\bibfnamefont {B.}~\bibnamefont {Ludoph}}, \bibinfo {author} {\bibfnamefont
  {A.}~\bibnamefont {Mart{\'\i}n-Rodero}}, \bibinfo {author} {\bibfnamefont
  {G.~R.}\ \bibnamefont {Bollinger}}, \bibinfo {author} {\bibfnamefont {J.~M.}\
  \bibnamefont {van Ruitenbeek}}, \ and\ \bibinfo {author} {\bibfnamefont
  {C.}~\bibnamefont {Urbina}},\ }\href@noop {} {\bibfield  {journal} {\bibinfo
  {journal} {Nature}\ }\textbf {\bibinfo {volume} {394}},\ \bibinfo {pages}
  {154} (\bibinfo {year} {1998})}\BibitemShut {NoStop}%
\bibitem [{\citenamefont {Cuevas}\ \emph {et~al.}(1998)\citenamefont {Cuevas},
  \citenamefont {Levy~Yeyati},\ and\ \citenamefont
  {Mart\'{\i}n-Rodero}}]{cuevas1998microscopic}%
  \BibitemOpen
  \bibfield  {author} {\bibinfo {author} {\bibfnamefont {J.~C.}\ \bibnamefont
  {Cuevas}}, \bibinfo {author} {\bibfnamefont {A.}~\bibnamefont {Levy~Yeyati}},
  \ and\ \bibinfo {author} {\bibfnamefont {A.}~\bibnamefont
  {Mart\'{\i}n-Rodero}},\ }\href@noop {} {\bibfield  {journal} {\bibinfo
  {journal} {Phys. Rev. Lett.}\ }\textbf {\bibinfo {volume} {80}},\ \bibinfo
  {pages} {1066} (\bibinfo {year} {1998})}\BibitemShut {NoStop}%
\bibitem [{\citenamefont {Senkpiel}\ \emph {et~al.}(2018)\citenamefont
  {Senkpiel}, \citenamefont {Dambach}, \citenamefont {Etzkorn}, \citenamefont
  {Drost}, \citenamefont {Padurariu}, \citenamefont {Kubala}, \citenamefont
  {Belzig}, \citenamefont {Levy~Yeyati}, \citenamefont {Cuevas}, \citenamefont
  {Ankerhold}, \citenamefont {Ast},\ and\ \citenamefont
  {Kern}}]{senkpiel2018single}%
  \BibitemOpen
  \bibfield  {author} {\bibinfo {author} {\bibfnamefont {J.}~\bibnamefont
  {Senkpiel}}, \bibinfo {author} {\bibfnamefont {S.}~\bibnamefont {Dambach}},
  \bibinfo {author} {\bibfnamefont {M.}~\bibnamefont {Etzkorn}}, \bibinfo
  {author} {\bibfnamefont {R.}~\bibnamefont {Drost}}, \bibinfo {author}
  {\bibfnamefont {C.}~\bibnamefont {Padurariu}}, \bibinfo {author}
  {\bibfnamefont {B.}~\bibnamefont {Kubala}}, \bibinfo {author} {\bibfnamefont
  {W.}~\bibnamefont {Belzig}}, \bibinfo {author} {\bibfnamefont
  {A.}~\bibnamefont {Levy~Yeyati}}, \bibinfo {author} {\bibfnamefont {J.~C.}\
  \bibnamefont {Cuevas}}, \bibinfo {author} {\bibfnamefont {J.}~\bibnamefont
  {Ankerhold}}, \bibinfo {author} {\bibfnamefont {C.~R.}\ \bibnamefont {Ast}},
  \ and\ \bibinfo {author} {\bibfnamefont {K.}~\bibnamefont {Kern}},\
  }\href@noop {} {\bibfield  {journal} {\bibinfo  {journal} {Arxiv pre-print}\
  ,\ \bibinfo {pages} {arXiv:1810.10609v1}} (\bibinfo {year}
  {2018})}\BibitemShut {NoStop}%
\bibitem [{\citenamefont {Cuevas}\ and\ \citenamefont
  {Belzig}(2003)}]{cuevas2003full}%
  \BibitemOpen
  \bibfield  {author} {\bibinfo {author} {\bibfnamefont {J.~C.}\ \bibnamefont
  {Cuevas}}\ and\ \bibinfo {author} {\bibfnamefont {W.}~\bibnamefont
  {Belzig}},\ }\href@noop {} {\bibfield  {journal} {\bibinfo  {journal} {Phys.
  Rev. Lett.}\ }\textbf {\bibinfo {volume} {91}},\ \bibinfo {pages} {187001}
  (\bibinfo {year} {2003})}\BibitemShut {NoStop}%
\bibitem [{\citenamefont {Johansson}\ \emph {et~al.}(2003)\citenamefont
  {Johansson}, \citenamefont {Samuelsson},\ and\ \citenamefont
  {Ingerman}}]{johansson2003full}%
  \BibitemOpen
  \bibfield  {author} {\bibinfo {author} {\bibfnamefont {G.}~\bibnamefont
  {Johansson}}, \bibinfo {author} {\bibfnamefont {P.}~\bibnamefont
  {Samuelsson}}, \ and\ \bibinfo {author} {\bibfnamefont {{\AA}.}~\bibnamefont
  {Ingerman}},\ }\href@noop {} {\bibfield  {journal} {\bibinfo  {journal}
  {Phys. Rev. Lett.}\ }\textbf {\bibinfo {volume} {91}},\ \bibinfo {pages}
  {187002} (\bibinfo {year} {2003})}\BibitemShut {NoStop}%
\bibitem [{\citenamefont {Cuevas}\ and\ \citenamefont
  {Belzig}(2004)}]{cuevas2004dc}%
  \BibitemOpen
  \bibfield  {author} {\bibinfo {author} {\bibfnamefont {J.~C.}\ \bibnamefont
  {Cuevas}}\ and\ \bibinfo {author} {\bibfnamefont {W.}~\bibnamefont
  {Belzig}},\ }\href@noop {} {\bibfield  {journal} {\bibinfo  {journal} {Phys.
  Rev. B}\ }\textbf {\bibinfo {volume} {70}},\ \bibinfo {pages} {214512}
  (\bibinfo {year} {2004})}\BibitemShut {NoStop}%
\bibitem [{\citenamefont {J{\"a}ck}\ \emph {et~al.}(2016)\citenamefont
  {J{\"a}ck}, \citenamefont {Eltschka}, \citenamefont {Assig}, \citenamefont
  {Etzkorn}, \citenamefont {Ast},\ and\ \citenamefont
  {Kern}}]{jack2016critical}%
  \BibitemOpen
  \bibfield  {author} {\bibinfo {author} {\bibfnamefont {B.}~\bibnamefont
  {J{\"a}ck}}, \bibinfo {author} {\bibfnamefont {M.}~\bibnamefont {Eltschka}},
  \bibinfo {author} {\bibfnamefont {M.}~\bibnamefont {Assig}}, \bibinfo
  {author} {\bibfnamefont {M.}~\bibnamefont {Etzkorn}}, \bibinfo {author}
  {\bibfnamefont {C.~R.}\ \bibnamefont {Ast}}, \ and\ \bibinfo {author}
  {\bibfnamefont {K.}~\bibnamefont {Kern}},\ }\href@noop {} {\bibfield
  {journal} {\bibinfo  {journal} {Phys. Rev. B}\ }\textbf {\bibinfo {volume}
  {93}},\ \bibinfo {pages} {020504} (\bibinfo {year} {2016})}\BibitemShut
  {NoStop}%
\end{thebibliography}

\begin{thebibliography}{8}
\expandafter\ifx\csname natexlab\endcsname\relax\def\natexlab#1{#1}\fi
\expandafter\ifx\csname bibnamefont\endcsname\relax
  \def\bibnamefont#1{#1}\fi
\expandafter\ifx\csname bibfnamefont\endcsname\relax
  \def\bibfnamefont#1{#1}\fi
\expandafter\ifx\csname citenamefont\endcsname\relax
  \def\citenamefont#1{#1}\fi
\expandafter\ifx\csname url\endcsname\relax
  \def\url#1{\texttt{#1}}\fi
\expandafter\ifx\csname urlprefix\endcsname\relax\def\urlprefix{URL }\fi
\providecommand{\bibinfo}[2]{#2}
\providecommand{\eprint}[2][]{\url{#2}}

\bibitem[{\citenamefont{Tinkham}(1996)}]{tinkham}
\bibinfo{author}{\bibfnamefont{M.}~\bibnamefont{Tinkham}},
  \emph{\bibinfo{title}{Introduction to Superconductivity}}
  (\bibinfo{publisher}{Dover Publications Inc.}, \bibinfo{year}{1996}),
  \bibinfo{edition}{2nd} ed.

\bibitem[{\citenamefont{Shapiro}(1963)}]{shapiro1963josephson}
\bibinfo{author}{\bibfnamefont{S.}~\bibnamefont{Shapiro}},
  \bibinfo{journal}{Phys. Rev. Lett.} \textbf{\bibinfo{volume}{11}},
  \bibinfo{pages}{80} (\bibinfo{year}{1963}).

\bibitem[{\citenamefont{Falci et~al.}(1991)\citenamefont{Falci, Bubanja, and
  Sch{\"o}n}}]{falci1991quasiparticle}
\bibinfo{author}{\bibfnamefont{G.}~\bibnamefont{Falci}},
  \bibinfo{author}{\bibfnamefont{V.}~\bibnamefont{Bubanja}}, \bibnamefont{and}
  \bibinfo{author}{\bibfnamefont{G.}~\bibnamefont{Sch{\"o}n}},
  \bibinfo{journal}{Z. Phys. B} \textbf{\bibinfo{volume}{85}},
  \bibinfo{pages}{451} (\bibinfo{year}{1991}).

\bibitem[{\citenamefont{Bratus et~al.}(1995)\citenamefont{Bratus, Shumeiko, and
  Wendin}}]{bratus1995theory}
\bibinfo{author}{\bibfnamefont{E.~N.} \bibnamefont{Bratus}},
  \bibinfo{author}{\bibfnamefont{V.~S.} \bibnamefont{Shumeiko}},
  \bibnamefont{and} \bibinfo{author}{\bibfnamefont{G.}~\bibnamefont{Wendin}},
  \bibinfo{journal}{Phy. Rev. Lett.} \textbf{\bibinfo{volume}{74}},
  \bibinfo{pages}{2110} (\bibinfo{year}{1995}).

\bibitem[{\citenamefont{Scheer et~al.}(1997)\citenamefont{Scheer, Joyez,
  Esteve, Urbina, and Devoret}}]{scheer1997conduction}
\bibinfo{author}{\bibfnamefont{E.}~\bibnamefont{Scheer}},
  \bibinfo{author}{\bibfnamefont{P.}~\bibnamefont{Joyez}},
  \bibinfo{author}{\bibfnamefont{D.}~\bibnamefont{Esteve}},
  \bibinfo{author}{\bibfnamefont{C.}~\bibnamefont{Urbina}}, \bibnamefont{and}
  \bibinfo{author}{\bibfnamefont{M.~H.} \bibnamefont{Devoret}},
  \bibinfo{journal}{Phys. Rev. Lett.} \textbf{\bibinfo{volume}{78}},
  \bibinfo{pages}{3535} (\bibinfo{year}{1997}).

\bibitem[{\citenamefont{Cuevas et~al.}(1996)\citenamefont{Cuevas,
  Mart{\'\i}n-Rodero, and Levy~Yeyati}}]{cuevas1996hamiltonian}
\bibinfo{author}{\bibfnamefont{J.~C.} \bibnamefont{Cuevas}},
  \bibinfo{author}{\bibfnamefont{A.}~\bibnamefont{Mart{\'\i}n-Rodero}},
  \bibnamefont{and}
  \bibinfo{author}{\bibfnamefont{A.}~\bibnamefont{Levy~Yeyati}},
  \bibinfo{journal}{Phys. Rev. B} \textbf{\bibinfo{volume}{54}},
  \bibinfo{pages}{7366} (\bibinfo{year}{1996}).

\bibitem[{\citenamefont{Cuevas et~al.}(1998)\citenamefont{Cuevas, Levy~Yeyati,
  and Mart\'{\i}n-Rodero}}]{cuevas1998microscopic}
\bibinfo{author}{\bibfnamefont{J.~C.} \bibnamefont{Cuevas}},
  \bibinfo{author}{\bibfnamefont{A.}~\bibnamefont{Levy~Yeyati}},
  \bibnamefont{and}
  \bibinfo{author}{\bibfnamefont{A.}~\bibnamefont{Mart\'{\i}n-Rodero}},
  \bibinfo{journal}{Phys. Rev. Lett.} \textbf{\bibinfo{volume}{80}},
  \bibinfo{pages}{1066} (\bibinfo{year}{1998}).

\bibitem[{\citenamefont{Scheer et~al.}(1998)\citenamefont{Scheer, Agra{\"\i}t,
  Cuevas, Levy~Yeyati, Ludoph, Mart{\'\i}n-Rodero, Bollinger, van Ruitenbeek,
  and Urbina}}]{scheer1998signature}
\bibinfo{author}{\bibfnamefont{E.}~\bibnamefont{Scheer}},
  \bibinfo{author}{\bibfnamefont{N.}~\bibnamefont{Agra{\"\i}t}},
  \bibinfo{author}{\bibfnamefont{J.~C.} \bibnamefont{Cuevas}},
  \bibinfo{author}{\bibfnamefont{A.}~\bibnamefont{Levy~Yeyati}},
  \bibinfo{author}{\bibfnamefont{B.}~\bibnamefont{Ludoph}},
  \bibinfo{author}{\bibfnamefont{A.}~\bibnamefont{Mart{\'\i}n-Rodero}},
  \bibinfo{author}{\bibfnamefont{G.~R.} \bibnamefont{Bollinger}},
  \bibinfo{author}{\bibfnamefont{J.~M.} \bibnamefont{van Ruitenbeek}},
  \bibnamefont{and} \bibinfo{author}{\bibfnamefont{C.}~\bibnamefont{Urbina}},
  \bibinfo{journal}{Nature} \textbf{\bibinfo{volume}{394}},
  \bibinfo{pages}{154} (\bibinfo{year}{1998}).

\end{thebibliography}


\end{document}